\documentclass[10pt]{article}

\usepackage{amsmath}

\usepackage{array}

\usepackage{appendix}

\usepackage{tocloft}                   

\usepackage{graphicx}

\usepackage{amsfonts}

\usepackage{amssymb}

\usepackage{mathrsfs}

\usepackage{yfonts}

\usepackage{euscript}

\usepackage{centernot}                 

\usepackage{ifsym}                     

\usepackage{upgreek}

\usepackage{mathtools}

\usepackage{bm}

\usepackage{color}

\usepackage{slantsc}
\usepackage{calligra}

\usepackage{bbold}          

\usepackage[T1]{fontenc}

\usepackage{epsf}

\usepackage{latexsym}

\usepackage{tipa}

\usepackage{makeidx}

\makeindex

\def\be{\begin{equation}}

\def\ee{\end{equation}}

\def\beq{\begin{equation}}

\def\eeq{\end{equation}}

\def\bea{\begin{eqnarray}}

\def\eea{\end{eqnarray}}

\def\!{\hspace{-1.6667em}}

\def\m{\mbox{ }}

\def\mma {\m , \m \m }

\def\!{\hspace{-1.6667em}}

\def\n{\noindent}

\def\u{\underline}

\def\w{\widetilde}

\def\es{\m = \m}

\def\:={\m := \m}

\def\=:{\m =: \m}

\def\bid{\mbox{\boldmath$d$}}

\def\bigg{\mbox{\boldmath$g$}}

\def\sbiq{\mbox{\scriptsize\boldmath$q$}}

\def\bip{\mbox{\boldmath$p$}}

\def\biq{\mbox{\boldmath$q$}}

\def\sbiq{\mbox{\scriptsize\boldmath$q$}}

\def\bir{\mbox{\boldmath$r$}}

\def\biG{\mbox{\boldmath \scriptsize $G$}} 

\def\biN{\mbox{\boldmath$N$}}

\def\biP{\mbox{\boldmath$P$}}

\def\biQ{\mbox{\boldmath$Q$}}

\def\biT{\mbox{\boldmath$T$}}

\def\biV{\mbox{\boldmath$V$}}

\def\sbiA{\mbox{\scriptsize\boldmath$A$}}

\def\sbiB{\mbox{\scriptsize\boldmath$B$}}

\def\sbiD{\mbox{\scriptsize\boldmath$D$}}

\def\sbiU{\mbox{\scriptsize\boldmath$U$}}

\def\sbiG{\mbox{\scriptsize\boldmath$G$}}

\def\sbiK{\mbox{\scriptsize\boldmath$K$}}

\def\sbiP{\mbox{\scriptsize\boldmath$P$}}

\def\sbiQ{\mbox{\scriptsize\boldmath$Q$}}

\def\sbiO{\mbox{\scriptsize\boldmath$O$}}

\def\sbiS{\mbox{\scriptsize\boldmath$S$}}

\def\sbiC{\mbox{\scriptsize\boldmath$C$}}

\def\bipi{\mbox{\boldmath$\pi$}}

\def\bnabla{\mbox{\boldmath$\nabla$}}               

\def\ma{\mbox{a}}

\def\me{\mbox{e}}

\def\mf{\mbox{f}}

\def\bp{\mbox{\bf p}}

\def\bsigma{\mbox{\boldmath$\sigma$}}                   %

\def\bupSigma{\mbox{\boldmath$\Sigma$}}                 

\def\sa{\mbox{\scriptsize a}}

\def\sU{\mbox{\scriptsize U}}

\def\sbcF{\mbox{\boldmath \scriptsize ${\cal F}$}}

\def\sbcG{\mbox{\boldmath \scriptsize ${\cal G}$}}

\def\tiG{\mbox{\tiny$K$}}

\def\tiC{\mbox{\tiny$C$}}

\def\tiD{\mbox{\tiny$D$}}

\def\sumi2{\sum\mbox{}_{\mbox{}_{\mbox{\scriptsize $i$=1}}}^2}

\def\sumi3{\sum\mbox{}_{\mbox{}_{\mbox{\scriptsize $i$=1}}}^3}

\def\sumABcycles3{\sum\mbox{}_{\mbox{}_{\mbox{\scriptsize cycles $A,B$=1}}}^{3}}

\def\sumCDcycles3{\sum\mbox{}_{\mbox{}_{\mbox{\scriptsize cycles $C,D$=1}}}^{3}}

\def\sumj3{\sum\mbox{}_{\mbox{}_{\mbox{\scriptsize $j$=1}}}^3}

\def\sumk3{\sum\mbox{}_{\mbox{}_{\mbox{\scriptsize $k$=1}}}^3}

\def\prodiA1{\prod\mbox{}_{\mbox{}_{\mbox{\scriptsize $i$=1}}}^{A - 1}}

\def\bigtimes{\mbox{\Large $\times$}}

\def\d{\textrm{d}}                                                  

\def\pa{\partial}                                                   

\def\siA{\mbox{\scriptsize $A$}}

\def\siB{\mbox{\scriptsize $B$}}

\def\tiO{\mbox{\tiny $O$}}

\def\tiU{\mbox{\tiny $U$}}

\def\tiD{\mbox{\tiny $D$}}

\def\tiG{\mbox{\tiny $G$}}

\def\tbiG{\mbox{\tiny\boldmath$G$}}

\def\tFrP{\mbox{\scriptsize$\mathfrak{P}$}}                               

\def\tFrQ{\mbox{\scriptsize$\mathfrak{Q}$}}                               

\def\FrS{\mbox{\Large $\mathfrak{s}$}}                         

\def\lFrg{\mbox{\Large$\mathfrak{g}$}}                         

\def\Hilb{\mbox{{\boldmath$\mathfrak{H}$}ilb}}                 

\def\scC{\mbox{\scriptsize ${\cal C}$}}                    

\def\bscC{\mbox{\scriptsize \boldmath${\cal C}$}}                    
  
\def\scE{\mbox{\scriptsize ${\cal E}$}}                    

\def\scH{\mbox{\scriptsize ${\cal H}$}}                    

\def\scM{\mbox{\scriptsize ${\cal M}$}}                    

\def\scP{\mbox{\scriptsize ${\cal P}$}}

\def\scS{\mbox{\scriptsize ${\cal S}$}}                    

\def\bFlin{\sbcF\mbox{\bf lin}} 

\def\Chronos{\scC\mbox{hronos}}                            

\def\bGauge{\sbcG\mbox{\bf auge}} 

\def\siD{\mbox{\scriptsize$D$}}                             %

\def\siO{\mbox{\scriptsize$O$}}                             %

\def\sbiU{\mbox{\scriptsize\boldmath$U$}}                             %

\def\sbiG{\mbox{\scriptsize\boldmath$G$}}                             %

\def\sbiO{\mbox{\scriptsize\boldmath$O$}}                             %

\def\tiG{\mbox{\tiny\boldmath$G$}}                                  
\def\FrQ{\mbox{\Large $\mathfrak{q}$}}                               
\def\tFrQ{\mbox{\scriptsize $\mathfrak{q}$}} 
												  
\def\bFrA{\mbox{\boldmath$\mathfrak{A}$}}                            
 	
\def\Phase{\mbox{{\boldmath$\mathfrak{P}$}hase}}                     
\def\tPhase{\mbox{\tiny{\boldmath$\mathfrak{P}$}hase}}         

\def\Mom{\mbox{{\boldmath$\mathfrak{P}$}}} 

\def\bFrR{\mbox{\boldmath$\mathfrak{R}$}}                            
\def\Rig-Phase{\bFrR\mbox{ig-}\Phase}                                
                                                                													   
\def\bFrM{\mbox{\boldmath${\mathfrak{M}}$}}                             

\def\bFrA{\mbox{\boldmath$\mathfrak{A}$}}                            
															   
\def\Positive-Modespace{\mbox{{\boldmath$\mathfrak{M}$}odespace$^+$}}

\def\POSITIVE-MODESPACE{\mbox{{\boldmath$\mathfrak{M}$}ODESPACE$^+$}}
                                                                                                                             														
\def\bFrA{\mbox{\boldmath$\mathfrak{A}$}} 		                     

\def\K{Kucha\v{r} }

\def\Kin-Hilb{\mbox{{\boldmath$\mathfrak{K}$}in-\Hilb}}                     

\def\Mid-Hilb{\mbox{{\boldmath$\mathfrak{M}$}id-\Hilb}}                     

\def\Dyn-Hilb{\mbox{{\boldmath$\mathfrak{D}$}yn-\Hilb}}                     

\def\5Star{\mbox{\Large$\star$}}              

\begin{document}

\begin{titlepage}

\begin{center}

\vspace{0.1in}

\Large{\bf Spaces of Observables from Solving PDEs} 

\vspace{0.1in}

\Large{\bf I. Translation-Invariant Theory.} \normalsize

\vspace{0.1in}

{\large \bf Edward Anderson$^*$}

\end{center}

\begin{abstract}

Finding classical canonical observables consists of taking a function space over phase space.
For constrained theories, these functions must form zero brackets with a closed algebraic structure of first-class constraints.
This brackets condition can moreover be recast as a first-order PDE system, to be treated as a free characteristic problem.
We explore explicit observables equations and their concrete solutions for a translation- and reparametrization-invariant action, 
thereby populating the following variety of notions of observables with examples.
1) The brackets can be strongly or weakly zero in Dirac's sense, i.e.\ a linear combination of constraints.  
2) Observables can admit pure-configuration and pure-momentum restrictions.
3) Our model provides the translation constraint ${\cal P}_i$ encoding zero total momentum of the model universe and depending homogeneous-linearly on momenta, 
and the ${\cal C}$hronos constraint equation of time reinterpretation of the `constant-energy condition' and depending quadratically on momenta.
These are mechanical analogues of GR's momentum  ${\cal M}_i$ and Hamiltonian ${\cal H}$ constraints respectively.  
${\cal P}_i$ and ${\cal C}$hronos moreover algebraically close separately (only ${\cal M}_i$ does for GR). 
Our model thus supports translation gauge-observables $G$ and Chronos observables $C$, 
as well as unrestricted observables $U$ and Dirac observables $D$ brackets-commuting with neither and both respectively. 
We relate the strong to properly-weak split of weak observables to the complementary-function to particular-integral split of the complete solution, 
with the properly-weak rendered of measure-0 relative to the strong by the latter's free characteristicness.  
The closed algebraic structures form a bounded lattice and the corresponding notions of observables a dual lattice, 
with the observables themselves forming a presheaf of function spaces thereover.

\end{abstract}
 
\n PACS: 04.20.Cv, 04.20.Fy, Physics keywords: observables, constraints. Background Independence. Problem of Observables facet of Problem of Time. 

\vspace{0.1in}

\n Mathematics keywords: Constrained systems. Geometrically-significant PDEs. Characteristic Problem. Integrability. Brackets algebras. Shape Theory. 

\vspace{0.1in}
  
\n $^*$ Dr.E.Anderson.Maths.Physics *at* protonmail.com 

\end{titlepage}

\section{Introduction}

\n The current Series of Articles \cite{DO-2, DO-3} concerns finding explicit examples of quite a number of 
distinct notions of observables \cite{DiracObs, HT92, Kuchar92, I93, Kuchar93, ABeables, ABook}. 
These arise in connection with Background Independence \cite{A64, A67, BB82, Giu06, APoT3, ASoS, ABook}, 
the Problem of Time \cite{BSW, WheelerGRT, Battelle, DeWitt67, Kuchar81, Kuchar91, Kuchar92, I93, Kuchar99, K04, R04, APoT, APoT2, ABook, PoT-Lett} 
(which can be viewed as difficulties arising in attempting to implement Background Independence)
and the study of constrained systems \cite{Dirac51, Dirac58, ADM, Dirac, HT92, SBook, ABook}.

\subsection{Classical canonical version of pure observables theory} 

\n In its purest form, moreover, {\it Assigning Observables} \cite{ABook} is a matter of {\it Taking Function Spaces Thereover} \cite{APoT3}.  
The current Series of Articles being in the {\it classical canonical} context, 
this means that {\it classical canonical observables} are assigned by finding functions\footnote{As the current Series" central conceptual class of objects,
observables are collectively picked out by undersized italic-letter notation.} 

\n\be 
\sbiO(\biP, \biQ)
\label{Obs}
\ee 
of {\it configurations} 

\n\be
\biQ
\ee 
and their {\it conjugate momenta} 

\n\be 
\biP            \m . 
\ee
Given a model

\n\be 
\FrS
\ee 
the totality of values taken by its $\biQ$ and $\biP$, as equipped by the {\it Poisson bracket}  

\n\be 
\mbox{\bf \{}  \m  \mbox{\bf ,}  \m  \mbox{\bf \}} \m , 
\ee 
constitutes {\it phase space} \cite{WoodhouseBook},  

\n\be 
\Phase(\FrS)    \m .   
\label{Phase}
\ee 
The space of the observables (\ref{Obs}), which we denote by 

\n\be 
{\cal O}(\FrS)  \m , 
\ee 
is a function space thereover.

\m 

\n We also consider the notion of {\it classical configurational observables} 

\n\be 
\sbiO(\biQ) = \sbiQ(\biQ)           \m ,
\label{o-p}
\ee 
forming a function space 

\n\be 
{\cal O}_{\sbiQ}(\FrS) = {\cal Q}    
\label{O-P}
\ee 
over the {\it configuration space} \cite{Lanczos, ABook, I, II, III}

\n\be 
\FrQ(\FrS)
\ee 
consisting of the totality of values taken by the given model $\FrS$'s configurations $\biQ$. 
(\ref{o-p}, \ref{O-P}) merit two distinct notations to emphasize that configurational observables $\sbiO(\biQ)$ in a physical context coincide moreover 
                                                               with {\it preserved quantities} 

\n\be
\sbiQ(\biQ)
\ee 
in a purely geometrical context \cite{PE-1, PE-2, PE-3}. 
This coincidence was recently derived in \cite{PE-1}, so the equality in (\ref{o-p}) has the status of a {\sl Theorem}.
This Theorem is moreover significant in {\sl bridging between} systematic ways of finding more involved notions of observables 
-- of both general and Background-Independence-specific use in Physics -- 
and systematic ways of finding preserved quantities for geometries: of both general and foundational use \cite{PE-1, PE-2, PE-3} in Geometry \cite{Hilb-Ax, HC32, Coxeter, Stillwell}.  
We now follow suit by using ${\cal O}_{\sbiQ}$ to denote {\it space of configurational observables} and ${\cal Q}$ to denote {\it space of geometrical preserved quantities},  
(\ref{O-P})'s equality between these having the status of a {\sl Corollary} of the above `Bridge Theorem'.  

\m 

\n We furthermore consider the notion of {\it classical momentum observables} 

\n\be 
\sbiO(\biP)  
\label{o-p-2}
\ee 
forming a function space 

\n\be 
{\cal O}_{\sbiP}(\FrS)                              \m , 
\label{O-P-2}
\ee 
over the {\it momentum space} 

\n\be 
\Mom(\FrS)  
\ee 
consisting of the totality of values taken by the given model $\FrS$'s momenta $\biP$. 

\m 

\n{\bf Remark 1} We require {\it suitably smooth} function spaces for the purposes of doing Classical Physics with the $\sbiO$ (and Differential Geometry with the $\sbiQ$); 
the current Series of Articles (and \cite{PE-1, PE-2, PE-3}) handles this by assuming 

\n\be 
{\cal C}^{\infty} \m \mbox{ functions}  \m . 
\ee 
\n{\bf Remark 2} Thus we have  

\n\be 
{\cal O}(\FrS)          \es  {\cal C}^{\infty}(\Phase(\FrS))                     \m , 
\ee 

\n\be 
{\cal O}_{\sbiQ}(\FrS)  \es  {\cal C}^{\infty}(\FrQ(\FrS)) = {\cal Q}(\FrS)      \m , 
\ee 

\n\be 
{\cal O}_{\sbiP}(\FrS)  \es  {\cal C}^{\infty}(\Mom(\FrS))                       \m .  
\ee
\n{\bf Remark 3} Configuration and momentum observables each readily represent a {\it restriction} of functions over $\Phase$ to functions just over $\FrQ$ and $\Mom$ respectively.   
These are, more specifically, {\it polarization restrictions} \cite{WoodhouseBook} since they precisely halve the number of variables 
(at least in the quadratic theories we consider in the current series, these being by far the simplest and most standard form for bosonic theories in Physics). 

\m 

\n{\bf Modelling Assumption 1} This precise halving makes particular sense in the context of finite quadratic models, for which $\FrQ$ is $k$-dimensional, 
$\Mom$ is then conjugately $k$-dimensional and $\Phase$ is $2 \, k$-dimensional.  

\m 

\n{\bf Modelling Assumption 2} We more specifically consider $N$ points (labelled $I = 1$ to $N$) on a given $d$-dimensional manifold 
\be 
\bFrM^d                                                                                                                   \m , 
\ee 
resulting in the {\it constellation space}

\n\be 
\FrQ(\bFrM^d, N)  \es  \bigtimes_{I = 1}^N \bFrM^d                                                                                 \m : 
\label{Constell}
\ee 
playing at least an incipient role as configuration space. 
In this setting, 

\n\be 
k := \mbox{dim}(\FrQ(\bFrM^d, N)) = N \, d                                                                                \m . 
\ee 
These notions apply both in the purely-geometrical context of $N$ points and in the physical context \cite{LR95, LR97, M02, M05, FileR, M15, Minimal-N} 
in which these $N$ points are materially-realized by $N$ particles: an $N$-Body Problem \cite{Marchal, LR97} (classical and nonrelativistic in the current Article's context).

\m 

\n{\bf Modelling Assumption 3} For simplicity, we additionally consider 

\n\be 
\bFrM^d = \mathbb{R}^d                                                                                                    \m ,
\ee 
for which 

\n\be 
\FrQ(d, N) := \FrQ(\mathbb{R}^d, N)  \es  \bigtimes_{I = 1}^N \mathbb{R}^d  \es \mathbb{R}^{N \, d}                            \m , 
\ee 

\n\be 
\Mom(d, N) := \Mom(\mathbb{R}^d, N)                               = \mathbb{R}^{N \, d}                                   \m , 
\ee 
and 

\n\be 
\Phase(d, N) := \Phase(\mathbb{R}^d, N)                           = \mathbb{R}^{2 \, N \, d}                              \m ,
\ee 
as equipped with the standard Poisson bracket 

\n\be 
\mbox{\bf \{} F \mbox{\bf ,}  \, G  \mbox{\bf \}}  \:=   \frac{\pa F}{\pa \biq}  \frac{\pa G}{\pa \bip} - \frac{\pa F}{\pa \bip}  \frac{\pa G}{ \pa \biq}
\ee 
for configurations $\biq$ ($\u{q}^I$ alias $q^{Ia}$ in components for underlines and $a$ indices denoting spatial vectors) 
and momenta        $\bip$ ($\u{p}_I$ alias $p_{Ia}$ in components).
Then 

\n\be
\sbiU  \es \sbiU(\biq, \, \bip)                                                                                             \m ,
\label{U(p,q)}
\ee 
denote {\it unrestricted observables}.
These form the {\it function space of unrestricted observables} 

\n\be 
{\cal O}_{\tiU}(\FrS)  \es  {\cal C}^{\infty}\left(  \mathbb{R}^{2 \, N \, d}  \right)                                  \m , 
\ee 
with restriction  -- configurational unrestricted observables --

\n\be
\sbiU(\biQ)  \es  \sbiU(\biq) 
            \es  \sbiQ                                                                                                     \m ,
\ee
forming 

\n\be 
{\cal O}_{\tiU(\sbiQ)}(\FrS)  \es  {\cal C}^{\infty}\left(  \mathbb{R}^{N \, d}  \right)  \es  {\cal Q}_{\tiU}(\FrS)      \m , 
\ee 
and restriction -- momentum unrestricted observables -- 

\n\be
\sbiU(\biP)  \es  \sbiU(\bip)                                                                                              \m ,
\ee
forming 

\n\be 
{\cal O}_{\tiU(\sbiP)}(\FrS)  \es  {\cal C}^{\infty}\left(  \mathbb{R}^{N \, d}  \right)                           \m .  
\ee
In this manner, the pure theory of (local classical canonical finite) observables is completed.

\subsection{How Observables Theory gains complexity}

\n A plethora of other fundamental considerations enter at this point, so providing an outline is adviseable and enlightening. 

\m 

\n{\bf Alternative 1} We could consider a pure theory of {\it spacetime} observables instead. 

\m 

\n{\bf Complexity 1} We could consider a pure theory of {\it quantum} observables. 

\m 

\n{\bf Complexity 2} We could consider a theory of {\it constrained} observables.  

\m 

\n{\bf Complexity 3} We could consider attaching further {\it global} and {\it topological} meaning to our theory of observables.  

\m 

\n{\bf Remark 1} Labelling canonical observables theory by Alternative 0), `1) or 0)' rests on what one is entertaining to be more primary: 
spacetime on the one hand, or one or more of space, configuration space or phase space on the other. 
Moreover, ab initio, the pure theory of spacetime observables is no more complicated than that of canonical observables. 
One just considers function spaces over one's model of spacetime,

\n\be
\bFrM  \m. 
\ee
\n{\bf Remark 2} Passing to quantum observables, however, represents a major surge in complexity. 
This is because only relatively few of the classical observables -- functions over phase space -- can be mutually-consistently promoted to quantum operators. 
One has to select a candidate subalgebra to promote, and then check that this manages to be consistent. 
But `selection' is in general highly ambiguous, and `promotion' as well 
(read `operator-ordering' \cite{DeWitt57, Magic, ABook}, `well-definedness' in the sense of `regularization' \cite{Weinberg1, Reg12} 
and `in possession of suitable properties to constitute a quantum operator' \cite{RS, KRBook}). 
And `consistency' both depends on all of the previous, and is rarely satisfied. 
That `subalgebra selection' and `consistency checks' are needed in the first place is due in good part to the {\it Groenewold--Van Hove Phenomenon} \cite{Gotay}, 
This has the added misfortune of being hard enough to deal with that the repertoire of specific examples for which this has been observed, and resolved when present, 
remains fairly small. 
Finally note that in this quantum case one is taking function spaces {\sl of operators} \cite{RS, KRBook} over {\sl quantum state space}.

\m 

\n{\bf Remark 3} While `constrained observables' is a canonical notion, spacetime has a somewhat analogous notion involving spacetime symmetry generators. 
In a nutshell,\footnote{Constraints are collectively picked out as a conceptual class of objects 
by being denoted by undersized calligraphic letters. 
Spacetime symmetry generators are denoted likewise, partly out of affinity, and partly though only featuring in the current subsection of this Series.} 
while the restriction 

\n\be 
\mbox{\bf |[} \, \scC \mbox{\bf ,} \, \sbiO \, \mbox{\bf ]|}  \m `=' \m  0 
\label{C-O-Gen}
\ee 
is required of canonical observables, an analogous restriction can be phrased for the spacetime version as well: 

\n\be 
\mbox{\bf |[} \, \scS \mbox{\bf ,} \, \sbiS \, \mbox{\bf ]|} \m  `=' \m  0 
\label{S-S-Gen}
\ee 
For now we use general notions of bracket 

\n\be 
\mbox{\bf |[} \, \m \mbox{\bf ,} \m \, \mbox{\bf ]|}
\ee 
and equality 

\n\be
`='
\ee 
(and constraint, where applicable), qualifying specific cases of what we mean later on.  

\m 

\n{\bf Example 1} For conventional Gauge Theory, (\ref{C-O-Gen}) imposes gauge invariance at the level of configuration space based on both space and internal gauge space. 
On the other hand, (\ref{C-O-Gen}) imposes gauge invariance at the more conventional level of spacetime and internal gauge space over this.

\m 

\n{\bf Example 2} The cause c\'{e}l$\grave{\me}$bre of canonical treatment of observables is GR.  
In this case, one has the spatial 3-diffeomorphisms 

\n\be 
Diff(\bupSigma)
\ee 
momentum constraint \cite{ADM}

\n\be
\scM_i
\ee 
which is linear in its momenta, and a Hamiltonian constraint \cite{ADM}

\n\be 
\scH
\ee 
-- quadratic in its momenta -- to commute with.

\m 

\n The spacetime version also has a GR case: commuting with the spacetime 4-diffeomorphisms' 

\n\be 
Diff(\bFrM)
\ee 
generators.  
Such restrictions can be hard to impose in practice \cite{Kuchar92, I93}. 

\m 

\n Finite models help understand some of the issues in hand. 

\m

\n{\bf Example 3}  {\it Minisuperspace} (spatially homogeneous GR) only has a $\scH$, and a single finite constraint oversimplifies the diversity of notions of observables. 

\m 

\n{\bf Example 4} Our finite mechanics models on $\mathbb{R}^n$ (\cite{BB82, FileR, AMech} and Sec 2), however, can be set up with momentum constraint analogues as well, 
so it is well-suited to the current Article's variety of notions of observables. 

\m 

\n Canonical-or-spacetime and unrestricted-or-constrained alternatives moreover clearly apply both at the classical and at the quantum level.  
It makes sense moreover to isolate study of constrained features in the absence of quantum before trying to work with both together 
[and vice versa, though we go for constrained but not quantum in the current Series]. 

\m 

\n We additionally sketch in Sec 2 how our model's constraints can be taken to be {\it provided} from two preceding Background Independence aspects;  
see Fig 1.a) for the historical evolution of the nomenclature of facet and aspect names.

\m 

\n{\bf Aspect 1} Temporal        Relationalism \cite{B94I, GLET, TRiPoD, ABook}, 
which gets one round the Frozen Formalism Problem Facet \cite{DeWitt67, Kuchar81, Kuchar92, I93, ABook} of the Problem of Time. 
This produces constraints quadratic in momenta, such as GR's $\scH$ or mechanics' `constant-energy condition' 

\n\be 
\scE = 0
\ee 
albeit reinterpreted as both a constraint and as an equation of time \cite{B94I, GLET}.
We thus denote $\scH$ and $\scH$'s conceptual class of constraint by 

\n\be 
\Chronos   \m . 
\ee 
\n{\bf Aspect 2} Configurational Relationalism \cite{BB82, FileR, ABook}, 
which gets one round the Best-Matching Problem Facet    \cite{BB82, FileR} of the Problem of Time (a generalization of the Thin Sandwich Problem \cite{WheelerGRT, BO, BF}). 
This produces first-class constraints which are linear in their momenta, including GR's $\scM_i$. 

\m 

\n{\bf Remark 1}  Conventional Gauge Theory implements Configurational Relationalism but not Temporal Relationalism.  

\m 

\n{\bf Remark 2}  While first-class constraints $\bFlin$ do no always coincide with gauge constraints $\bGauge$ \cite{HT92, ABook}, 
this coincidence does hold within the current Series' range of examples, so we treat

\n\be 
\bFlin = \bGauge 
\ee
interchangeably. 

\m 

\n{\bf Aspect 3} Constraint Closure is then required, 
by which a Dirac-type Algorithm \cite{Dirac51, Dirac, HT92, ABook} is evoked to assess if these constraints provided close in a mutually-compatible manner.  

\m 

\n The main lesson of the Problem of Time \cite{Kuchar92, I93, APoT} is moreover that the lion's share of the resolution is in compatibilizing single-facet strategies to jointly apply.
It is thus quite a considerable issue for {\it Assigning Observables} -- that at the classical level it can only be faced as {\bf Aspect 4} of Background Independence. 
I.e.\ after the two constraint providers and Constraint Closure's check on the mutual consistency of these constraints, 
as per the (3 facets, then 1 facet) decoupling outlined in the next subsection.  

\m 

\n The other local aspects-or-facets (Fig 1) moreover do not further interfere with this canonical observables facet.  
For now, what we note is that Complexity 2) is already a 2-facet interplay. 
This historically caused difficulties by being taken as the starting point for classical observables theory, thus missing Sec 1.1's pure form entirely.\footnote{Historical 
and pedagogical difficulties with observables moreover abound, with many quantum treatises and popular accounts \cite{Obs-W} immediately launching into the quantum version instead, 
and not even mentioning classical precursors or further difficulties due to constraints.  
It is thus important to clarify that observables are both a classical and a quantum issue, 
                                                     both a canonical and a spacetime issue, with substantial further complication from constraints or symmetry generators, 
                                                 and both a local and a global issue.
It furthermore helps considerably to point out that what one is doing in each case is Taking Function Spaces Thereover: an apparent triviality in the 
unconstrained classical local case that becomes highly nontrivial in diverse ways upon passing to whichever combination of constrained, quantum and global.}

{            \begin{figure}[!ht]
\centering
\includegraphics[width=0.75\textwidth]{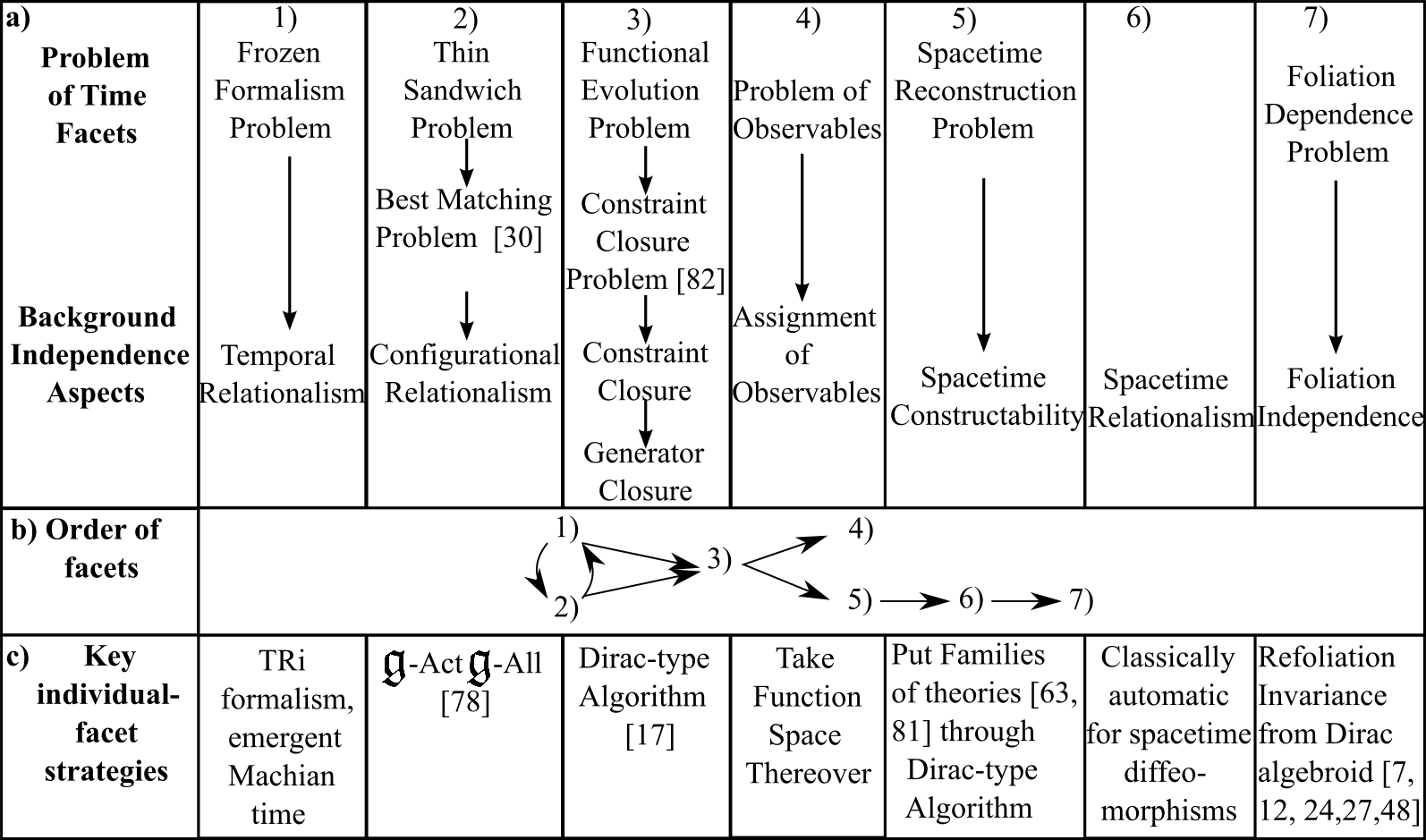}
\caption[Text der im Bilderverzeichnis auftaucht]{\footnotesize{a) Evolution of conceptualization and nomenclature of Problem of Time facets. 
The first row are Kucha\v{r} and Isham's classification from the 1990's \cite{Kuchar92, I93}, 
though 'frozen', `thin sandwich', and the need for spacetime construction date back to Wheeler and DeWitt in the 1960s \cite{WheelerGRT, DeWitt67, Battelle} 
and the involvement of 4- and 3-diffeomorphisms even further back. 
This Figure moreover links these facet names and concepts to Background Independence \cite{A64, A67, Giu06} aspects \cite{APoT3, ABook}, 
from the position that difficulties with implementing Background Independence aspects {\sl result in} Problem of Time facets.
All bar the last of these aspects are already classically present. 
b)  The order in which the facets are incorporated, which itself resolves a longstanding problem \cite{Kuchar92, I93, Kuchar93}.
c) The key piecemeal strategies subsequently combined to make a local resolution of the Problem of Time in \cite{ABook, PoT-Lett}. 
We indicate their origins in 1950s and 60s work of Dirac, 1970s work of Teitelboim and Kucha\v{r}, 1990's work of Isham and 1980s-00s work of Barbour.
Aspects 5 to 7 are not realized in the current Series' mechanics models; they also do not interfere with assignment of canonical observables procedure even when present \cite{ABook}.
Down Aspect 6's the spacetime route, one Background Independence aspect is a {\it generator} provider leading to its own consideration of {\it Generator Closure}.
Facet 8 are Multiple-Choice Problems, which is purely quantum mechanical and principally concerns the Groenewold--Van Hove Phenomenon; Facet 9 are the Global Problems of Time.} }
\label{Evol-Fac}\end{figure}            }
 
\subsection{Constraint Consistency precedes Assigning Observables}

For now we just suppose we have some constraints, and make some algebraic inferences.

\m 
 
\n In the current Series, we consider sets of classical constraints which moreover close as Poisson brackets algebras:  

\n\be
\mbox{\bf \{} \scC_A \mbox{\bf ,} \, \scC_B \mbox{\bf \}}  \es  {C^C}_{AB} \, \scC_C             \m , 
\label{C-C}
\ee
for ${C^C}_{AB}$ the corresponding {\it constraint algebra structure constants}.

\m 

\n Classical brackets-closing defines first-class constraints. 
We do not encounter any second-class constraints in the current series; these do not brackets-close, but can be eliminated by the formation of a further classical bracket.
I.e.\ the Dirac bracket \cite{Dirac, Sni, HT92}, which then assumes the role of classical bracket (hence `classical brackets close' rather than just `Poisson brackets close').
See Complexity $0^{\prime}$ is Sec 9.2 for another subtlety arising at this point.

\m 

\n We also require the following distinction between weak and strong vanishing \cite{Dirac}.
On the one hand, `weak' is meant in the sense of Dirac \cite{Dirac}, meaning zero `up to a homogeneous-linear combination of the constraints'. 
On the other hand, `strong' just means identically zero.  

\m 

\n So for the purpose of the current series, our bracket is Poisson and our `=' is strong = or weak $\approx$; 
this distinction applies {\sl again} at the level of defining observables \cite{DiracObs, HT92, ABeables}.

\m 

\n Classical canonical observables equations moreover take the form of a {\sl system} of brackets equations, 
\n\be
\mbox{\bf \{} \scC_A \mbox{\bf ,} \, \sbiO \mbox{\bf \}}  \m `=' \m 0             \m .  
\label{O-BE}
\ee
Let us distinguish between {\it strong observables} solving
\n\be
\mbox{\bf \{} \scC_A \mbox{\bf ,} \, \sbiO \mbox{\bf \}}  \es     0             \m .  
\label{S-O-BE}
\ee
and {\it weak observables} solving  
\n\be
\mbox{\bf \{} \scC_A \mbox{\bf ,} \, \siO_P \mbox{\bf \}}  \es     {W^{B}}_{AP} \, \scC_B             \m ,   
\label{W-O-BE}
\ee
for {\it weak linear dependence structure constants} ${W^B}_{AP}$.
Let us furthermore define {\it properly-weak observables} by exclusion of strong observables from weak observables. 

\m 

\n{\bf Theorem 1} Only sets of constraints $\bscC$ which form closed algebraic stuctures form a valid system of observables equations. 

\m 

\n This consistency follows from Jacobi's identity acting on two copies of $\bscC$ and one of $\sbiO$: 

\n\be 
             \mbox{\bf{|[}} \,\mbox{\bf{|[}}  \scC_A \mbox{\bf ,} \, \scC_B \mbox{\bf ]|,} \, \sbiO  \mbox{\bf ]|}                     \es  
- \left\{ \, \mbox{\bf{|[}} \,\mbox{\bf{|[}}  \scC_B \mbox{\bf ,} \, \sbiO  \mbox{\bf ]|,} \, \scC_A \mbox{\bf ]|}  \m + \m  
             \mbox{\bf{|[}} \,\mbox{\bf{|[}}  \sbiO  \mbox{\bf ,} \, \scC_A \mbox{\bf ]|,} \, \scC_B \mbox{\bf ]|}  \,  \right\} 
																			                                                        \m `=' \m  0  \m + \m  0 
																			                                                        \m `=' \m        0        \m .
\ee 
Thus $\sbiO$ must also form `zero' general brackets with $\mbox{\bf |[} \scC_A \mbox{\bf ,} \, \scC_B \mbox{\bf ]|}$. 
So the quantities forming `zero' brackets brackets-close.    
Thus general brackets closure of constraints is crucial {\it prior to} defining notions of observables.  

\m 

\n Observables carry furthermore an internal index of their own: $\siO_O$.   
This has not entered almost any equations up to this point because all the $\siO_O$ solve the observables equation regardless 
($\siO_O$ is a $\scC_A = \scC_{a \, g}$ scalar: a spatial scalar and a scalar with respect to the original constraint algebraic structure). 
We mention this internal index at this point due to its entering the algebraic theory of the preserved quantities and the observables themselves.

\m 

\n{\bf Theorem 2} Observables themselves must close as an algebra, 

\n\be
\mbox{\bf |[} \siO_O \mbox{\bf ,} \, \siO_P \mbox{\bf ]|}  \es  {O^Q}_{OP} \, \siO_Q           \m , 
\label{O-O}
\ee
where ${O^Q}_{OP}$ are the corresponding {\it observables algebra structure constants}.

\m 

\n This follows from Jacobi's identity acting on two copies of $\sbiO$ and one of $\bscC$: 

\n\be 
\mbox{\bf{|[}} \,\mbox{\bf{|[}} \siO_O \mbox{\bf ,}  \, \siO_P \mbox{\bf ]|,} \, \scC_A \mbox{\bf ]|}                                  \es 
- \left\{ \,  
\mbox{\bf{|[}} \,\mbox{\bf{|[}} \siO_P \mbox{\bf ,}  \, \scC_A \mbox{\bf ]|,} \, \siO_O \mbox{\bf ]|}  \m + \m 
\mbox{\bf{|[}} \,\mbox{\bf{|[}} \scC_A \mbox{\bf ,}  \, \siO_O \mbox{\bf ]|,} \, \siO_P \mbox{\bf ]|}           \, \right\} 
                                                                                                                             \m `=' \m  0 \m + \m 0 
																					   			                             \m `=' \m       0               \m . 
\ee 
So $\scC_A$ must also form `zero' general bracket with $\mbox{\bf|[}\siO_O \mbox{\bf,}\, \siO_P \mbox{\bf]|}$, by which the latter itself obeys the defining property of the $\sbiO$. 
I.e.\ the $\sbiO$ are algebraically closed under general brackets. 

\m 

\n Theorem 1 moreover applies to any consistent algebraic structure.
The entire bounded lattice of constraint subalgebraic structures of a given model $\FrS$ can thus be used to 
define a dual bounded lattice's worth of notions of observables (see Fig 2).   
Each of these is an observables algebra by Theorem 2 applying. 
%
{            \begin{figure}[!ht]
\centering
\includegraphics[width=1.0\textwidth]{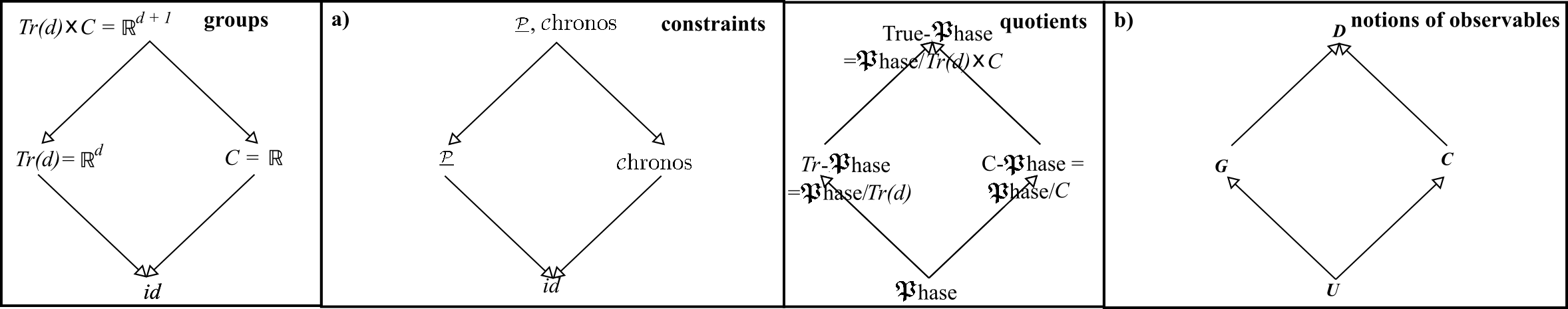}
\caption[Text der im Bilderverzeichnis auftaucht]{        \footnotesize{a) Bounded lattice of constraints, with its underlying bounded lattice of subgroups 
and its dual lattice of phase space quotients by those subgroups. b) The corresponding dual bounded lattice of notions of observables.}}
\label{Sim-Latt} \end{figure}          }

\m

\n On the one hand, the bottom notion of a closed algebra of constraints is no constraints, 
to which unrestricted observables are a dual-bottom notion i.e.\ a top notion of observable. 

\m

\n On the other hand, the top notion is all the first-class constraints $\sbcF$, to which {\it Dirac observables} $\sbiD$ obeying 
\be 
\mbox{\bf \{} \sbcF \mbox{\bf ,} \, \sbiD \mbox{\bf \}}  \m `=' \m 0                             \m .  
\ee
are a dual-top notion i.e.\ a bottom notion of observable. 

\m 

\n For some physical theories, the first-class linear constraints $\bFlin$ moreover close algebraically, 
by which these support the corresponding notion of Kucha\v{r} observables \cite{Kuchar93}, $\sbiK$  
\be 
\mbox{\bf \{} \bFlin \mbox{\bf ,} \, \sbiK \mbox{\bf \}} \m `=' \m 0   \m . 
\ee 
The lattice of notions of observables is moreover a theory-independent generalization of the possibility of there being some kinds of `middling' observables, 
a role played in GR by the Kucha\v{r} observables themselves. 
 
\m  

\n First-class linear constraints $\bFlin$ and gauge constraints $\bGauge$ are moreover not always exactly the same, by which Kucha\v{r} observables $\sbiK$  
can be distinct from {\it gauge observables} $\sbiG$ commuting with the latter, 
\be 
\mbox{\bf \{} \bGauge \mbox{\bf ,} \, \sbiG \mbox{\bf \}} \m `=' \m 0  \m .
\ee
This distinction does not enter the current Series' geometrical examples, and so is left to other treatises \cite{HT92, ABook}.
In summary throughout the current series' specific examples, 

\n\be  
\sbiK = \sbiG  \m . 
\ee
\n{\bf Example 1} For GR, the $\bFlin = \scM_i$ close by themselves, but $\scH$ does not.  Thus GR supports, $\sbiU$, $\sbiK$ and $\sbiD$.  

\m 

\n{\bf Example 2} The current series' Mechanics models have an analogue of $\scH$ -- the $\Chronos$ reinterpretation of the `constant energy condition' $\scE$, 
which moreover closes by itself.
These models thus support a notion of {\it Chronos observables} $\sbiC$ obeying 
\be 
\mbox{\bf \{} \Chronos \mbox{\bf ,} \, \sbiC \mbox{\bf \}} \m `=' \m 0  \m .
\ee
\n{\bf Remark 1} The unrestricted observables are the easiest to find, whereas the Dirac observables are the hardest to find, out of being the most restricted.

\m 

\n{\bf Remark 2} All of the above notions of observables all come in strong and weak versions; 
this is a doubling in some places of what function spaces can be attached to our lattice of notions (or more than doubling, if one considers finer distinctions like $\lFrg$-weak)

\m 

\n For the corresponding function spaces of observables, we use plain
\n\be
{\cal O}                                        \mma  
\ee 
in the case of strong observables, 

\n\be 
{\cal O}^w                                      \mma 
\ee 
for properly weak observables,  and 

\n\be
{\cal O}^W  \es {\cal O} \m \coprod  \m {\cal O}^w
\ee 
for the whole spaces of weak observables. 
(This disjoint union is in accord with properly-weak being defined by exclusion of the strong from among the weak.)
We furthermore hang $\sbiU$, $\sbiG$, $\sbiK$, $\sbiC$, $\sbiD$ (...) suffices on these spaces as suits.      

\m 

\n{\bf Remark 3} Some of these notions admit their own restrictions to configurational and momentum notions of observables. 
For $\lFrg$-notions of constraints and observables, 
configurational observables remain the same as preserved quantities corresponding to the geometry with automorphism group $\lFrg$.

\m

\n{\bf Remark 4} There is also at least ab initio an indirect-reduced-relational trichotomy.
This is between handling Configurational Relationalism indirectly and not reducing out the ensuing constraints, reducing them out, and avoiding having any in the first place by 
handling Configurational Relationalism ab initio in gauge-invariant variables.
Each of these is in principle capable of distinct realization of some notions of observables.  

\m
	
\n{\bf Remark 5} Such a large zoo of conceptually meaningful and technically sharply-defined notions of observables does well to be populated with simple concrete examples. 
The main function of Secs 3 to 8 of the current Article, and of Articles 2 and 3, is to provide a start on this.

\subsection{PDE reformulation}

\n We can find such examples by reformulating the observables brackets equations (\ref{S-O-BE},\ref{W-O-BE}) as specific PDE systems \cite{ABeables, ABook, PE-1}.   
Namely, 
\be 
\sum_{I = 1}^N \, \left\{ \frac{\pa \scC_A}{\pa q^{Ia}} \frac{\pa \siO_P}{\pa p_{Ia}} - \frac{\pa \scC_A}{\pa p_{Ia}} \frac{\pa \siO_P}{\pa q^{Ia}} \right\} \es 0 
\ee 
in the strong case, and 
\be 
\sum_{I = 1}^N \, \left\{ \frac{\pa \scC_A}{\pa q^{Ia}} \frac{\pa \siO_P}{\pa p_{Ia}} - \frac{\pa \scC_A}{\pa p_{Ia}} \frac{\pa \siO_P}{\pa q^{Ia}} \right\} \es   {W^B}_{AP}  \,\, \scC_B   
\ee 
in the weak case.     
\cite{PE-1} moreover classified these PDEs, argued for them to be treated as Free Characteristic Problems, and provided an array of specific solving techniques.
In a nutshell, 

\m 

\n I) observables equations are first-order systems of $\Delta = 1$ to $D$ PDEs for $\sbiO$, 
\be 
F_{\Delta}(\biq, \sbiO, \pa_{\sbiq}\sbiO)  \es  0 \m .
\ee 
\n II) Observables equations are furthermore homogeneous-linear in the strong case 

\n\be 
\sum_{aI = 1}^{N \, d} {A^{aI}}_{\Delta}(\biq)\pa_{q^{aI}}\sbiO  \es  0 \mma 
\ee 
and with additional inhomogeneous terms in the weak case, 

\n\be 
\sum_{aI = 1}^{N \, d} {A^{aI}}_{\Delta}(\biq)\pa_{q^{aI}}\sbiO  \es  b_{\Delta}(\biq) \mma 
\ee 
Preserved quantities do not admit moreover admit a weak extension\cite{PE-1}, 
so observables' versatility in this regard is one of the ways in which the current series expands on \cite{PE-1, PE-2, PE-3}.   

\m 

\n III) In the homogeneous case, constants solve, but are termed the trivial solution and are excluded from the thus-defined {\it proper observables}. 
%

\m 

\n IV) {\it Proper observables in the second sense} is the brackets-theoretically-formulable exclusion of observables that are purely functions of constraints, 
since these obviously commute with a closed constraint algebraic structure and are moreover but `pure gauge fluff' rather than physical observables.    

\m 

\n V) The single-equation system of the above (whether homogeneous or inhomogeneous) is equivalent to an ODE system by the flow methods; 
the Characteristic Problem formulation for this is then a standard prescription \cite{CH1, John}. 
For resulting characteristic surface 
\be 
\chi \m , 
\ee
\be 
{\cal O} = \{ {\cal C}^{\infty} \m \mbox{ functions on } \m \chi  \mma  \mf(\chi) \} \m .  
\ee 
\n VI) In adopting more specifically a {\sl Free} Characteristic Problem -- whether for a single observables PDE or a system of them --
we comment firstly that `free', alias `natural' \cite{CH1} signifies specifically treating the general problem rather than subjecting it to prescribed data.   
Secondly, that this is both geometrically and physically appropriate as suits presecribed quantities and observables respectively.  
Thirdly, that it is {\sl the very embodiment} of Taking Function Spaces Thereover. 
 
\m 

\n VII) The system version is not however a standard prescription.  
It is crucial here that there is only one unknown function, so the over-determined possibility is realized. 

\m 

\n VIII) Such generically have no (proper) solutions, but the underlying constraint algebra closing (Theorem 1) comes to the rescue again, 
now by guaranteeing \cite{PE-1} integrability a la Frobenius \cite{AMP, Lee2} for the observables PDE system. 

\m 

\n IX) In the current Series' specific models, translational constraints are amenable to being removed 
(at least within the affine family of geometries or of Relational Mechanics thereupon), by passing to the centre of mass. 
Upon doing this, Jacobi coordinates \cite{Marchal} furthermore furnish a widespread simplification of one's remaining equations \cite{FileR, I, Minimal-N}.   
This relates ab-initio-translationless and translation-reduced actions, constraints, Hamiltonians and observables, as being of the same form but with one object less 
(see Sec 2 for details).   

\m 

\n X) Further Articles in this series shall outline and extend more of \cite{PE-1, PE-2, PE-3}'s preserved-or-observables PDE system solving methodology, 
as suits solving for furtherly complicated examples.  

\m 

\n XI) New to the current article, we tie strong, properly-weak and weak notions of observables 
to the complementary function to particular integral split of complete solutions of PDEs. 

\m 

\n XII) The Free Characteristic Problem posed in VI) moreover leads to consideration of intersections of characteristic surfaces, 
which can moreover be conceived of in terms of restriction maps. 
I.e.\ if 
\be 
\sbiA \m \mbox{ such that } \m \mbox{\bf [} \scC_A \mbox{\bf ,} \, \sbiA \mbox{\bf ]} = 0 \m \mbox{ forms characteristic surface } \m \chi_{A}  \mma 
\ee 
\be 
\sbiB \m \mbox{ such that } \m \mbox{\bf [} \scC_B \mbox{\bf ,} \, \sbiB \mbox{\bf ]} = 0 \m \mbox{ forms characteristic surface } \m \chi_{B}  \mma  
\ee 
then 
\be 
\sbiO \m \mbox{ such that } \m \mbox{\bf [} \scC_C \mbox{\bf ,} \, \sbiB \mbox{\bf ]} = 0 \mma C = A \, \cup \,  B 
\mbox{ forms characteristic surface } \m \chi_{C}  =  \chi_{A} \, \cap \chi_{B} 
                                                   = \chi_{A}\mbox{\Large$|$}_{_{\mbox{\bf [} \scC_{\siB} \mbox{\bf ,} \, \sbiA \mbox{\bf ]} \mbox{\scriptsize = 0}}} 
												   = \chi_{B}\mbox{\Large$|$}_{_{\mbox{\bf [} \scC_{\siA} \mbox{\bf ,} \, \sbiB \mbox{\bf ]} \mbox{\scriptsize = 0}}}  \m . 
\ee 

\subsection{Outline of rest of the paper}

\n Sec 2 introduces our translation-invariant geometrical mechanics model, which gives a simple model of the first four Background Independence aspects/ Problem of Time facets.
We subsequently find the strong and then weak $Tr$-gauge, Chronos and Dirac observables in Secs 3 to 8 respectively, including the function spaces formed in each case.  
Our Conclusion Sec 9 finally motivates a further range of instructive examples, building on the introductory account below.

\subsection{Concerning future work}

\n Companion Articles 2 and 3 extend this program to, respectively, 
Kendall's preshapes [quotienting out dilatations $Dilatat(d)$:  dilations $Dil$ as well as translations]  
      and shapes    [quotienting out similarities $Sim(d)$ : rotations $Rot(d)$ as well as translations and dilations]. 
These give somewhat more elaborate -- and iconoclastic --model of the first four Background Independence aspects/ Problem of Time facets.
Indeed, the latter includes the most common relational side of the Absolute versus Relational Motion debate as a subcase: 
quotienting out the Euclidean group of translations and rotations. 

\m 

\n The Conclusion also points to this Program's capacity for extending of the Foundations of Geometry \cite{Hilb-Ax, HC32, Coxeter, S04, Stillwell} via preserved quantities.  
It also points to upcoming work on the field-theoretical counterpart, in which first-order {\sl functional} DE systems occur

\m 

\n Let us finally mention Complexity 3) in connection with the further future of Observables Theory. 
This global complexity applies moreover to {\sl all} previously mentioned combinations of complexitities. 
At a more advanced level, our `Taking Function Spaces Thereover' position on observables can be conceived of in terms of finding {\sl presheaves} of functions.
These are multiple function spaces over a given space which are moreover inter-related by restriction maps [c.f.\ item XII) in this regard]. 
So our position becomes `Taking Function Spaces Thereover including Restriction Maps Therebetween', i.e.\ {\it Taking Presheaves Thereover}.
There is moreover some hope to further transcend to Taking Sheaves Thereover, these carrying furthermore powerful global-gluing and local-recovery structures. 
So our position would become `Taking Function Spaces Thereover including Restriction Maps Therebetween, Gluing and Localization', i.e.\ `Taking Sheaves Thereover'.    
See in particular \cite{Ghrist, Wells, Wedhorn} for introductory accounts of sheaves and \cite{Sheaves1, Sheaves2, Hartshorne} for more advanced accounts, 
or Fredehagen and Haag's \cite{FH87} for a pioneering paper on the use of presheaves to model quantum observables.  

\m 

\n While a treatment of observables as presheaves is somewhat more advanced than the current series of articles, each Article's Conclusion does summarize that Article's findings 
in presheaf form, over the bounded lattice of notions of observables. 
The corresponding figure suffices to furthermore identify both of these as contravariant presheaves: 
their restriction maps' arrows run in opposition to the underlying lattice's ordering arrows. 

\m 

\n {\sl Whether} observables can be modelled by sheaves is a highly interesting question for the future development of Theoretical, Foundational and Mathematical Physics.   

\vspace{10in}

\section{Our model and its constraints}

\subsection{Indirect formulation}

\n Our model action is the Jacobi-type \cite{Lanczos} action
\be 
S = \int \d s \sqrt{2 \, W}  \mma 
\label{S}
\ee 
for $W = E - V$ and kinetic arc element 
\be 
\d s = \sqrt{   \delta_{ab}\delta_{IJ}\d q_{\sa}^{aI}\d q_{\sa}^{bJ}  }  \m ,
\ee    
where $\d \biq_{\sa}$ is the translation-corrected change, 
\be 
\d \biq_{\sa} = \d \biq - \d \u{\ma}  \m .
\ee 
{\bf Temporal Relationalism} \cite{FileR, APoT3, ABook} -- Leibnizian a priori timelessness for the Universe as a whole -- is implemented by (\ref{S}) 
by being homogenous-linear in its change variables. 
See \cite{ABook} for completion of this treatment of Temporal Relationalism with emergent Machian time \cite{M, Clemence, B94I, GLET}.

\m 

\n{\bf Configurational Relationalism} \cite{FileR, APoT3, ABook} -- physical irrelevance of a group of transformations $\lFrg$ -- 
is implemented by (\ref{S}) in the particular case of the group of translations, 
\be 
\lFrg = Tr(d) = \mathbb{R}^d  \m .
\label{Tr}
\ee
Our model is a for now a simplified version of Euclidean Relational Mechanics \cite{BB82, FORD, FileR}, in the sense of omitting corrections with respect to rotations. 
For $d = 1$ (the continuous part of) 
\be 
Eucl(1) = Tr(1)       \m ,
\ee 
so in this case, translational observables are Euclidean observables as well.  
Corrections with respect to dilations are more straightforward to implement \cite{Kendall84, Kendall}, so we next include these in Article 2, 
postponing  rotational corrections to Article 3, with \cite{BB82, LR97, FileR} and without \cite{Kendall84, Kendall, B03, FORD, FileR} dilational corrections. 

\m 

\n In greater generality, for a manifold $\bFrM$ equipped with a level of mathematical structure $\bsigma$, 
\be
\langle \bFrM, \, \bsigma \rangle \m , 
\label{Msig}
\ee 
{\sl Shape Theory} \cite{Kendall84, Kendall89, Small, Kendall, Bhatta, DM16, PE16, I, II, III, Minimal-N} considers not only the constellation space (\ref{Constell}) 
but also the effect of quotienting this by the automorphism group corresponding to (\ref{Msig}), with 
\be 
\FrS(\bFrM, \, \bsigma, N)  \:=  \frac{ \bigtimes_{I = 1}^N \, \langle \bFrM , \, \bsigma \rangle }{ Aut(\bFrM, \bsigma) }
\ee 
defining the corresponding {\sl shape space}.  

\m 

\n{\bf Example 1} If $\bsigma = \bigg$ is metric structure, for which 
\be 
Aut(\bFrM, \, \bigg ) = Isom( \bFrM, \, \bigg )            \m : 
\ee
the isometry group. 

\m 

\n{\bf Example 2} If $\bsigma = \bar{\bigg}$ is similarity structure -- metric structure modulo constant rescaling -- for which 
\be 
Aut(\bFrM, \, \bar{\bigg }) = Sim(\bFrM, \, \bar{\bigg} )  \m : 
\ee
the similarity group.  
One can moreover `count out' to establish the minimal $N$ for a given $\bFrM$ and $Aut(\bFrM, \, \bsigma)$ \cite{Minimal-N, Minimal-N-2}, using 
\be 
\mbox{dim}(\FrQ(\bFrM, \, \bsigma))  \m - \m \mbox{dim}(Active\mbox{$-$}Aut( \bFrM, \, \bsigma)) \m \geq \m m
\ee
for `Active-' denoting the active portion of the automorphism group on the configuration and $m$ the smallest number of degress of freedom allowed by one's study.  

\m 

\n Sec 1.2 moreover places Temporal and Configurational Relationalism within a large picture, as the first two of the nine conceptual aspects of Background Independence, 
difficulties with the implementation of which constitute the Problem of Time.  

\m  

\n Our action has the following implications.

\m 

\n On the one hand, variation with respect to the Configurational Relationalism implementing auxiliary variables $\u{\ma}$ \cite{FEP} provides, as a secondary constraint vector 
\be 
\u{\scP}  \es  \sum_{I = 1}^N \u{p}_I  \es  0            \m .  
\ee
\n On the other hand, the Temporal Relationalism implementing homogeneous-linearity of $\d \biq$ gives, 
by a variant \cite{FileR, ABook} of a basic argument of Dirac \cite{Dirac} as a primary constraint,  
\be 
\Chronos  \es  \frac{\bip^2}{2} - W  
          \es  \frac{1}{2} \, \sum_{I = 1}^N |\u{p}_I|^2 - W          \m .  
\ee 
This is computationally the same as a `fixed energy condition', but is now to be interpreted as an equation of time \cite{B94I, GLET, ABook}. 

\m  

\n{\bf Constraint Closure} -- the third aspect of Background Independence -- holds moreover as follows. 
On the one hand, $\u{\scP}$ self-closes as the commutative Lie algebra 
\be 
\mbox{\bf \{} \scP_a\mbox{\bf ,} \, \scP_b \mbox{\bf \}} \es  0       \m , 
\label{P-P}
\ee 
corrsponding to a realization of the underlying translation group (\ref{Tr}).

\m 

\n On the other hand, being a single finite scalar constraint, $\Chronos$ trivially self-closes: 
\be 
\mbox{\bf \{} \Chronos \mbox{\bf ,} \, \Chronos \mbox{\bf \}}  \es  0                     \m .  
\label{Ch-Ch}
\ee 
\n While the above constraints are provided by two distinct mechanisms each tied to a distinct aspect of Background Independence,  
they are moreover compatible, as is established by their mutual bracket closure, 

\n\be 
\mbox{\bf \{} \scC_A \mbox{\bf ,} \, \siO_P \mbox{\bf \}} \approx 0 \m \Rightarrow \m 
\mbox{\bf \{} \scC_A \mbox{\bf ,} \, \siO_P \mbox{\bf \}}   \es    {M^B}_{AP} \scC_B  
\ee
for {\it mutual structure constants} ${M^B}_{AP}$, is required. 

\m 

In the current series' case of constant $W$, this closure is moreover in the specific form 
\be 
\mbox{\bf \{} \scP_a \mbox{\bf ,} \, \Chronos \mbox{\bf \}}  =  0                       \m \Rightarrow  \m {M^B}_{AP} = 0  \m :  
\label{P-Ch}
\ee
strong closure.  

\m 

\n The above closure (\ref{P-P}, \ref{Ch-Ch}, \ref{P-Ch}) furthermore establishes that $\scP_A$ and $\Chronos$ are first-class, 
by which \cite{Dirac} their closure under the incipient Poisson bracket as used above is definitive.
It amounts to succeeding in establishing classical-level Constraint Closure -- a third aspect of Background Independence -- 
by completion of a simple case of a \cite{TRiPoD, AM13, ABook} Dirac-type Algorithm \cite{Dirac, HT92}.  
Together, $\scP_a$ and $\Chronos$ form the $(d + 1)$-dimesional noncompact commutative algebra
\be 
\bFrA = \mathbb{R}^{d + 1}                                                         \m .  
\ee 
See Sec 9.2 for more general and yet consistent choices of potential.
Inspection of the above three brackets reveals moreover that both $\u{\scP}$ and $\Chronos$ constitute closed constraint subalgebras, hence Fig 2.a). 

\m

\n Thus in turn our model supports Fig 2.b)'s dual lattice of $\sbiU$, $\sbiG$, $\sbiC$ and $\sbiD$ notions of observables, 
the $\sbiG$ being equivalent to $\sbiK$ in the current Series' range of models.  
These correspond to commuting with no constraints, just $\u{\scP}$, just $\Chronos$, and both constraints respectively.

\subsection{Reduction}

\n One can moreover reduce the translational constraint $\u{\scP}$ out, or never have a such in the first place in the directly-implemented relational formulation.

\m 

\n 1) A more straightforward way of attaining this is at the {\it Hamiltonian level} -- i.e.\ in 
\be 
( \biq , \, \bip ) 
\ee
variables -- by just substituting $\u{\scP}$ into $\Chronos$. 

\m 

\n 2) A less straightforward, but also instructive way, is at the {\it Jacobi--Mach level} -- i.e.\ in 
\be 
( \biq , \, \d \biq , \, \d\u{\ma} ) 
\ee 
variables -- by solving $\u{\scP}$ for $\d \u{\ma}$ and substituting this into the action. 
This is done moreover according to dRouthian reduction. 

\m 

\n To explain 2), consider first the analogous working in {\it Lagrangian variables} -- 

\n\be 
\left( \biq \mma   \frac{\d \biq}{\d\lambda}  \mma  \frac{\d \u{\ma}}{\d\lambda} \right)
\ee 
-- for $\lambda$ a meaningless parameter.
The familiar Routhian reduction applies here as regards elimination of cyclic velocities, a role played in our model by the auxiliary $\d \u{\ma}/\d \lambda$. 
{\it dRouthian redution} is then the analogous elimination of cyclic differentials $\d \u{\ma}$ from Jacobi--Mach variables actions. 
We use this rather than Routhian reduction so as to continue maintaining Temporal Relationalism, this explaining also our use of Jacobi actions, 
differential $\lFrg$ corrections, \cite{FEP}'s variational procedure, and our choice of version of Dirac-type Algortithm. 
It is by such perserverances and consistencies that the Author recently succeeded in providing a local resolution of the Problem of Time, meaning not only finding 
a way round each local facet of it, but also the much harder task \cite{Kuchar92, I93} of succeeding in rendering each of these resolutions compatible with the others. 
1) is moreover itself Temporal Relationalism compatible, by judicious choice of a slightly modified version of the Hamiltonian formulation; 
this requires less explanation than 1) due to standard Hamiltonians being {\sl closer} to attaining Temporal Relationalism than Lagrangians are.  

\m 

\n In 2), the end product of dRouthian reduction is a reduced action.  
For $N = 2$, this is readily fomulated in terms of $r$. 
For $N \geq 3$, one can continue to use $r$ variables but they render the reduced action non-diagonal, obscuring its simple structure. 
This is remedied by passing to {\it mass-weighted relative Jacobi coordinates}
\be 
\rho^i = \sqrt{\mu_i}R^i                                          \m . 
\label{MW-JC}
\ee
On the one hand, the $R_i$ here are relative Jacobi coordinates, 
\be 
R_1 = q_1 - q_2 = r_{12}                                          \m ,
\ee 
\be 
R_2 = q_1 + q_2 - 2 \, q_3                                        \m ,   
\label{R2}
\ee 
for $N = 3$.
Bases of relative Jacobi coordinates are not uniquely determined in various ways \cite{FileR, I, Minimal-N} but any choice thereof will do for the purposes of the current Series.  

\m 

\n On the other hand, the $\mu_i$ are the corresponding Jacobi masses $\mu_i$ (corresponding particle clusters' reduced masses), 
\be 
\mu_1 = \frac{1}{2} \mma \mu_2 = \frac{2}{3} 
\ee 
for $N = 3$.  
The $\rho^i$ are functionally a diagonalization, and conceptually a generalization from interparticle separations to interparticle cluster separations \cite{Marchal, I, Minimal-N}.  
Let us finally also denote the conjugate mass weighted relative Jacobi momenta by $\pi_i$.

\m 

\n Note that this looks just like the unconstrained observables version but with one object less. 
This is the Jacobi map: simply losing one object upon passing to the centre of mass frame (discarding the centre of mass's own absolute position).  

\n\be 
\sum_{I = 1}^N f(\u{q}^I) \m \longrightarrow \m \sum_i^n f(\u{\rho}^i)          \m .  
\label{J-Map}
\ee 
This works in the context of the current series of articles because the Jacobi coordinates diagonalize the configuration space metric $M_{abIJ}$.  
This points to $\mathbb{R}^{Nd}$ being sent to $\mathbb{R}^{nd}$: relative space is just a Cartesian space with $d$ coordinates less than the incipient constellation space.
This renders the current Article very geometrically straightforward, and systematic in many ways over all $(d, N)$.
Article 2 continues this geometrical simplicity and universal $(d, \, N)$ application: 
quotienting out $Dil$ sends one to $\mathbb{S}^{Nd - 1}$ or $\mathbb{S}^{nd - 1}$ spheres, the latter being Kendall's preshape sphere \cite{Kendall84, Kendall}. 
In 1-$d$ Preshape space is shape space. 
In 2-$d$ shape space is also universally tractable in simple geometrical terms by the generalized Hopf map sending one to $\mathbb{CP}^{N - 1}$. 
$d \geq 3$ is harder, though Article 3's specific examples involving rotations lie within the 2-$d$ confines of the above simple and systematical geometrical treatment.

\subsection{Directly-formulated relative action}

\n One can also arrive at an action from relative-space first principles. 
\cite{FileR} moreover showed that direct formulation and reduction coincide in this case. 
This in fact applies to Article 2's preshape space and shape space actions as well. 

\m 

\n We celebrate by calling it the {\it r-action}, standing for reduced-or-relative: 
\be 
\w{S}_r = \int \d \w{s} \sqrt{W(\w{Q})} \mma 
\ee
for r-kinetic line element 
\be
\d s = \w{M}_{AB} \d \w{Q}^A \d \w{Q}^B   \m .  
\ee 
\n This r-action contains no auxiliaries (but attains $Tr$-invariance by distinct direct means). 
But this now produces no $Tr$-constraint.  

\m 

\n The Dirac-type argument for a primary constraint giving an r-Chronos equation moreover continues to apply, giving 
\be 
\w{\Chronos} \:=  \frac{1}{2}|\pi_i|^2 - W  
             \es  0                                                           \m 
\ee 
(r-objects are often picked out using the tilde notation). 

\m 

\n Being a single finite constraint, this furthermore trivially self-closes: 
\be 
\mbox{\bf \{} \w{\Chronos} \mbox{\bf ,} \,  \w{\Chronos} \mbox{\bf \}} \es 0  \m .
\ee 
This case thus yields constraint algebras: $\Chronos$ by itself and no constraints ($id$). 

\m 

\n Correspondingly, the r-model supports just $\w{\sbiU}$ and $\w{\sbiD}$ as notions of observables.

\m 

\n The current Article shows moroever that these are essentially just $\sbiG$ and $\sbiD$.

\section{Strong \biT\bir-observables}

For models with gauge group of physically meaningless transformations $Tr(d)$, 
gauge $\sbiG$ and Kucha\v{r} $\sbiK$ observables coincide in the form of {\it translational observables}.

\subsection{r-approach}

In this case, there are no constraints.
The Jacobi map (\ref{J-Map}) furthermore gives the identification 
\be 
\w{\sbiG}(N, \, d) = \sbiU(n, \, d)  \m . 
\ee 
Thus from (\ref{U(p,q)})
\be 
\w{\sbiG}  \es  \w{\sbiG}(\u{\rho}^I, \u{\pi}_I)   
           \es  \w{\sbiG}({\bm{-}}, \, {\bm{-}}_{\sbiP})  \m , 
\label{41}
\ee 
using \cite{AMech, PE-1}'s shorthand ${\bm{-}}$ for differences of positions \cite{AMech, PE-1} and the new shorthand $ {\bm{-}}_{\sbiP}$ for differences of momenta.
It is moreover implicit that the $\sbiG$ are suitably-smooth, taken in the current series to mean ${\cal C}^{\infty}$.  

\m

\n These form the {\it function space of strong reduced Tr-observables}  
\be 
{\cal O}_{\w{\tbiG}}  \es  {\cal C}^{\infty}(\mathbb{R}^{2 \, n \, d})                             \m .
\label{42}
\ee 
This furthermore immediately restricts 

\m 

\n i) to strong configuration $Tr$-observables 
\be 
\w{\sbiG}(\biQ)  \es  \w{\sbiG}(\u{\rho}^I)  
                 \es  \w{\sbiG}({\bm{-}}) 
				 \es  \sbiQ_{\tbiG}  \m ,  
\ee 
forming the {\it function space of strong reduced configuration Tr-observables}  
\be 
{\cal O}_{\w{\tbiG}(\sbiQ)}(d, \, N, \, Tr)  \es  {\cal C}^{\infty}(\mathbb{R}^{n \, d})  
                                \es  {\cal Q}(d, \, N, \, Tr)                                   \m .  
\ee
ii) To strong momentum $Tr$-observables 
\be 
\w{\biG}(\biP)   \es  \w{\sbiG}(\u{\rho}^I)  
                  \es  \w{\sbiG}({\bm{-}}_{\sbiP})                                        \m ,
\ee 
forming the {\it function space of strong reduced momentum $Tr$-observables}   
\be 
{\cal O}_{\w{\tbiG}(\sbiP)}  \es  {\cal C}^{\infty}(\mathbb{R}^{n \, d})                             \m . 
\ee 
{\bf Remark 1} $N = 1$ -- i.e.\ $n = 0$ -- collapses, so $N = 2$ is minimal to nontrivially realize these three function spaces.

\subsection{Indirect approach}

In the indirect approach, the corresponding {\it strong 1-}$d$ {\it Euclidean} or {\it translational gauge observables brackets equation} is 
\be 
\mbox{\bf \{} \scP_i \mbox{\bf ,} \, \sbiG \mbox{\bf \}}  \es  0  \m .
\label{Tr-Ind-PDE}
\ee 
This implies the {\it translational gauge observables PDE system}

\n \be 
\sum_{I = 1}^N \u{\nabla}_{q^I} \sbiG  \es  0                \m . 
\label{TOE}
\ee 
Below we shall exploit that 
\be
\mbox{(\ref{TOE}) coincides with one of \cite{PE-1}'s preserved equations modulo our version carrying extra \m $p_I$ \m dependence.}
\label{PE-Trick}
\ee
\n Let us also introduce the small-model notation 
\be 
x := q_1 \mma y := q_2 \mma z := q_3                  \m .
\ee

\subsection{(\bid, \biN) = (1, 1)}

In this case, our PDE system (\ref{TOE}) collapses to 
\be 
\pa_x \sbiG = 0                   \m : 
\label{1}
\ee
a single first order homogeneous-linear PDE that is immediately integrable to  
\be 
\sbiG = \sbiG(p)  \m . 
\ee
But also for $N = 1$, 
\be 
p = \scP      \m , 
\ee 
so our solution is not proper in the second sense. 
Thus $N = 1$ admits no gauge observables: 
\be 
{\cal O}_{\tiG}(1, \, 1, \, Tr)  \es  \emptyset        \m . 
\ee
By restrictions of $\emptyset$ can just be $\emptyset$ again,
\be 
\left. \emptyset \right|_{_{\sU}} = \emptyset \m , 
\label{Res-Empty}
\ee 
it is clear that this model has no room for configuration or momentum versions of observables either, 
\be 
{\cal O}_{\tiG(\sbiP)}(1, \, 1, \, Tr)  \es  \emptyset   
                                        \es  {\cal O}_{\tiG(\sbiQ)}(1, \, 1, \, Tr)  
										\es  {\cal Q}_{\tiG}(1, \, 1, \, Tr) \m .  
\ee
\n{\bf Remark 1} Having to cut out improper solutions is a new feature relative to \cite{PE-1, PE-2, PE-3}; this reflects that in the context of constrained systems, 
first-class constraints use up two degrees of freedom each \cite{Dirac}, whereas preserved equations for geometries-or-configuration spaces use up only one. 
This in turn rests on phase space having double the number of degrees of freedom than configuration space by possessing one momentum degree of freedom per configurational one.

\subsection{(\bid, \biN) = (1, 2)}

Our PDE system (\ref{TOE}) now collapses to 
\be 
( \pa_x + \pa_y ) \sbiG = 0           \m : 
\label{2-1}
\ee
a single first order homogeneous-linear PDE in two variables.

\m 

\n (\ref{PE-Trick}) gives its solution as     
\be  
\sbiG  \es  \sbiG(x - y , \, p_x , \, p_y)  \m .   
\ee
\be 
p_x + p_y = \scP 
\ee
dependence is not however proper. 
We address this by a second trick, 
\be 
\mbox{If functions of a basis $x_i$ solve, so do arbitrary linear combinations $y^i = {M^i}_j x^j$ for $\u{\u{M}}$ invertible} \m ,  
\label{Trick-2}
\ee
by which the change of variables 
\be
P_{\pm} = p_x \pm p_y                
\ee 
isolates the proper solution to be  
\be  
\sbiG  \es  \sbiG(x - y , \, p_x - p_y) \es \sbiG(r, p_r)    \m . 
\ee
The $r$ variable here stands for `relative separation'.    

\m 

\n These form the {\it function space of strong Tr-observables} 
\be 
{\cal O}_{\tiG}(1, \, 2 , \, Tr)  \es  {\cal C}^{\infty}(\mathbb{R}^2)             \m , 
\ee
the $\mathbb{R}^2$ being realized by the characteristic functions $r$ and $p_r$.  

\m  

\n Aside from being the smallest model supporting strong $Tr$-observables, (1, 2) separately supports, firstly, strong configuration $Tr$-observables, 

\n\be 
\sbiG(\biQ)  \es  \sbiG(r)    \m ,
\ee 
forming 

\n\be 
{\cal O}_{\tiG(\sbiQ)}(1, \, 2 , \, Tr)  \es  {\cal C}^{\infty}(\mathbb{R})      \es  {\cal Q}_{\tiG}(1, \, 2 , \, Tr)       \m . 
\ee
Secondly, strong momentum $Tr$-observables

\n\be 
\sbiG(\biP)  \es  \sbiG(p_r)   \m ,
\ee 
forming 
\n\be 
{\cal O}_{\tiG(\sbiP)}(1, \, 2 , \, Tr)  \es  {\cal C}^{\infty}(\mathbb{R})  \m . 
\ee

\subsection{(\bid, \biN) = (1, 3)}

The strong gauge observables equation is now 
\be 
( \pa_x + \pa_y + \pa_z ) \sbiG = 0                               \m : 
\label{3-1}
\ee

\m 

\n Thus the solution is  
\be  
\sbiG  \es  \sbiG(x - z , \, y - z , \, p_x , \, p_y, \, p_z)           \m . 
\ee
Moreover, 
\be 
p_x + p_y + p_z  =  \scP 
\ee
dependence is not proper. 
Thus by the change of variables 
\be
P_0 = p_x + p_y + p_z                                                 \m , 
\ee
\be 
P_1 = p_x - p_z                                                       \m ,
\ee
\be 
P_2 = p_y - p_z                                                       \m ,
\ee
the proper solution is 
\be  
\sbiG  \es  \sbiG( x - z , \, y - z , \, p_x - p_z , \, p_y - p_z )     \m . 
\ee
Taking linear combinations (\ref{Trick-2}) now in the mass-weighted relative Jacobi form (\ref{MW-JC}-\ref{R2}) gives the rearrangement 
\be 
\sbiG  \es  \sbiG(\rho_1, \rho_2, \pi_1, \pi_2)                                     \m ,   
\ee 
which furthermore usefully generalizes to all larger $(d, N)$.  
Note that this looks just like the unconstrained observables case but with one object less; indeed, it is another example of the Jacobi map (\ref{J-Map}). 

\m 

\n These observables constitute  
\be 
{\cal O}_{\tiG}(1, \, 3, \, Tr)  \es  {\cal C}^{\infty}(\mathbb{R}^4)         \m , 
\ee 
the $\mathbb{R}^4$ being realized by the 4 characteristic functions $x - z$, $y - z$, $p_x - p_z$, $p_y - p_z$, or their linear combinations $\rho^i$, $\pi_i$, $i = 1$ to 2.

\subsection{The general case}

The general-$(d, N)$ case is now immediate.
The translational observables equation (\ref{TOE}) is solved by 
\be 
\sbiG  \es  \sbiG \left( \u{q}^i - \u{q}^N , \, \u{p}_i - \u{p}_N \right)  
       \es  \sbiG \left( \u{\rho}^i , \, \u{\pi}_i \right)                                            \m .  
\label{51}
\ee 
These form 
\be 
{\cal O}_{\tiG}(d, \, N, \, Tr)  \es  {\cal C}^{\infty}(\mathbb{R}^{2 \, n \, d})        \m , 
\label{52}
\ee 
the $\mathbb{R}^{2 \, n \, d}$ being realized by the $2 \, n \ d$ characteristic functions $q^i - q^N$, $p_i - p_N$, or their linear combinations $\rho^i$, $\pi_i$.  
If we seek configuration observables,    
\be  
\sbiG(\biQ)  \es  \sbiG \left( \u{q}^i - \u{q}^N \right) 
             \es  \sbiG \left( \u{\rho}^i \right) 
			 \es  \sbiG({\bm{-}})
\ee 
survives, realizing 
\be 
{\cal Q}(d, \, N, \, Tr)  \es  {\cal O}_{\tiG(\sbiQ)}(d, \, N, \, Tr)  
                    \es  {\cal C}^{\infty}(\mathbb{R}^{n \, d})                          \m . 
\ee 
On the other hand, if we seek momentum observables $\sbiG(\biP)$,  
\be  
\sbiG  \es  \sbiG \left( \u{p}_i - \u{p}_N \right) 
       \es  \sbiG \left( \u{\pi}_i \right)
       \es  \sbiG({\bm{-}}_{\sbiP})	   
\ee
survives, giving a distinct realization of 
\be 
{\cal O}_{\tiG(\sbiP)}(d, \, N, \, Tr)  \es  {\cal C}^{\infty}(\mathbb{R}^{n \, d})        \m . 
\ee 
\n{\bf Remark 1} By equations (\ref{41}, \ref{42}) match (\ref{51}, \ref{52}), the r- and indirectly-formulated notions of strong $Tr$-observables coincide.

\section{Weak \biT\bir-observables}

This is only distinct from the preceding section's working in the indirectly formulated case, since the reduced case has no room for any properly weak notion.  

\m 

\n The {\it weak 1-}$d$ {\it Euclidean} or {\it translational gauge observables equation} is 

\n\be 
- \sum_{I = 1}^N \u{\nabla}_{q^I} \sbiG^W  \es  k \, \u{\scP}  
                                           \es  k \, \sum_{I = 1}^N p_I      \m ,  
\label{WGE}
\ee 
This is an inhomogeneous extension of the preceding Sec's PDE, with $k \in \mathbb{R}_*$ supporting properly-weak observables.
The preceding section can moreover be reinterpreted as finding the complementary function (C.F.) for this, 
so only finding the particular integral (P.I.) part of the solution remains.

\subsection{(\bid, \biN) = (1, 1)}

\n In this case, our PDE system (\ref{WGE}) collapses to   
\be 
- \pa_x \sbiG^W = k \, p  = k \, {\scP}                                       \m . 
\label{2}
\ee
This supports the P.I.  
\be 
\sbiG^w = - k \, p \,  x \m . 
\label{PWO-11}
\ee
This is not solely a function of constraints, so it is proper.
It is additionally a smooth function. 
Since we showed this case admits no C.F.\, this is moreover the complete solution, $\sbiG^W$.  
I.e.\ for this model, the properly-weak observables (\ref{PWO-11}) also encompass the entirety of the weak observables.  

\m 

\n The corresponding {\it function space of properly-weak $Tr$-observables} is parametrized by $k$, so  
\be 
{\cal O}^w_{\tiG}(1, \, 1, \, Tr)  \es  \mathbb{R}_*  \m . 
\ee
For (1, 1), this coincides with the {\it function space of weak $Tr$-observables}
\be 
{\cal O}^W_{\tiG}(1, \, 1, \, Tr) \es  \mathbb{R}_*   \m . 
\ee
\n{\bf Remark 1} The appended properly-weak sector is moreover non-generic: of measure-zero relative to nontrivial spaces of strong observables.

\m 
 
\n{\bf Remark 2} Our observables moreover do not qualify as either configuration or as momentum observables due to their necessarily-mixed dependence on $x$ and $p$.  
Thus 
\be 
{\cal O}^W_{\tiG(\sbiP)}(1, \, 1, \, Tr)  \es   {\cal O}^w_{\tiG(\sbiP)}(1, \, 1, \, Tr)  
                                          \es   \emptyset 
                                          \es   {\cal O}^w_{\tiG(\sbiQ)}(1, \, 1, \, Tr)               
										  \es   {\cal O}^W_{\tiG(\sbiQ)}(1, \, 1, \, Tr)                    \m . 
\ee 
\n{\bf Remark 3} In fact, properly-weak configurational observables are never possible, since we are in a context \cite{PE-1} in which constraints have to depend on momenta
\be 
\scC = f(\biQ, \, \biP) \m \mbox{ \sl with specific $\biP$ dependence } \m .   
\ee 
By this, the weak configurational observables equation would have a momentum-dependent inhomogeneous term right-hand side. 
But this is inconsistent with admitting a solution with purely $\biQ$-dependent right-hand-side., unless both sides are separately zero. 
But the left-hand-side being zero returns us to the strong configuration observables equation.

\subsection{(\bid, \biN) = (1, 2)}\label{Sym}

\n In this case, our PDE system (\ref{WGE}) collapses to  
\be 
- ( \pa_x + \pa_y ) \sbiG^w  \es  k  \, \scP  
                           \es  k ( p_x + p_y )  \m . 
\label{2-2}
\ee
This supports the properly-weak P.I. 
\be 
\sbiG^w  \es - ( a \, x + b \, y ) \, ( p_x + p_y ) \mma a + b = k \mma 
\label{Sym-PI}
\ee 
which, by the preceding section is also the complete solution for the weak observables   

\m 

\n{\u{Derivation}}
Our PDE is equivalent by the flow method to the ODE system 
\be 
\dot{x} =  1   \m , 
\ee
\be 
\dot{y} =  1   \m , 
\ee
\be 
\dot{p}_x = 0  \m , 
\ee
\be 
\dot{p}_y = 0  \m , 
\ee
\be 
\dot{\sbiG} = - k ( p_x + p_y )                                                                                                          \m , 
\ee
to be treated as a Free Characteristic Problem.  
Integrating, 
\be 
x = t + u   \m , 
\ee 
\be 
y = t       \m , 
\ee
\be 
p_x = v     \m , 
\ee
\be 
p_y = w     \m , 
\ee
\be 
\sbiG \es \sbiG(u, v, w) \m - \m  \alpha( v + w )t \m + \m \beta                                                                                        \m , 
\ee
Eliminating $t$ between the first two of these equations, 
\be 
u = x - y \m . 
\ee
Thus, also striking the constant $\beta$ out for not being proper,  
\be 
\sbiG  \es  - k ( p_x + p_y ) y   \m + \m  \sbiG(x - y, \, p_x - p_y)                                                                    \m .  
\ee 
One can just as well attach $u$ to $y$, in which case $x$ enters this solution's first term. 
Treating $x$ and $y$ symmetrically, the P.I. part  
\be 
\sbiG = - k \, \frac{x + y}{2} ( p_x + p_y )
\ee 
follows, as does (\ref{Sym-PI}) by taking the arbitrary linear combination.                                                                $\Box$

\m 

\n $k \, \in \mathbb{R}_*$ continues to apply, while $a$ is a free real number: $a \in \mathbb{R}$.   
Thus
\be 
{\cal O}^w_{\tiG(\sbiQ)}(1, \, 2, \, Tr)  \es  \mathbb{R} \times \mathbb{R}_*       \m . 
\ee

\subsection{The general case}

For arbitrary $(N, \, d)$, the properly-weak $Tr$-observables take the form  

\n\be 
\sbiG  \es  - \sum_{a = 1}^{d} \sum_{I = 1}^N \lambda_{Ia} q^{Ia} \sum_{J = 1}^N p_{Ja}    \m + \m \sbiQ\left(\u{q}^{i} - \u{q}^{N}\right)  \mma \sum_{I = 1}^N \lambda_{Ia} = k       \m , 
\ee
and constitute 
\be 
{\cal O}^w_{\tiG(\sbiQ)}(N, \, d, \, Tr)  \es  \mathbb{R}^{n \, d} \times \mathbb{R}_*                                                                                      \m . 
\ee
\n {\bf Remark 4} Remarks 1 and 2 extend to arbitrary $(d, N)$, so the space of properly weak observables is of measure zero relative to that of strong observables, 
and no configuration or momentum weak $Tr$-observables are supported.

\section{Strong Chronos observables for \biV = const}

\subsection{The strong Chronos observables equation}

\n This follows from our model possessing a $\Chronos$ constraint reinterpretation of an energy equation;  
this moreover only makes sense as an indirect formulation.
In this case, notions of Chronos observables are supported.  
For now, the {\it strong Chronos observables brackets equation} is 
\be 
\mbox{\bf \{} \Chronos \mbox{\bf ,} \, \sbiC \mbox{\bf \}} \es     0  \m .
\label{C-Ind-PDE}
\ee
This implies the {\it strong Chronos observables PDE}

\n\be 
\bip \circ \bnabla \, \sbiC  \es  \sum_{I = 1}^N p_I \pa_{q^I} \sbiC  
                             \es  0                                                            \m . 
\label{SCE}
\ee

\subsection{(\bid, \biN) = (1, 1)}

In this case, our PDE (\ref{SCE}) simplifies to
\be 
p \, \pa_x \sbiC = 0                                                                                  \m ,   
\ee
which factorizes into the following. 

\m 

\n i) A PDE (\ref{1}), thus supporting 
\be 
\sbiC = \sbiC(p)                                                                                       \m . 
\ee 
But 
\be 
p = \sqrt{2(\Chronos + W)}
\ee
is a function of the constraint $\Chronos$ alone for $V$ and thus $W$ constant, so, once again, these are not proper observables. 

\m 

\n ii) An algebraic equation 
\be 
p = 0                                                                                                \m ,   
\label{p = 0}
\ee 
so 
\be 
\sbiC = \sbiC(x)                                                                                       \m . 
\ee 
(\ref{p = 0}) is however clearly but a static, i.e.\ non-dynamical possibility, so we do not further entertain it.   
By this reckoning, 
\be 
{\cal O}_{\tiC}(1, \, 1, \, Tr)  \m = \m \emptyset                                                   \m . 
\ee

\subsection{(\bid, \biN) = (1, 2)}\label{SC-12}

Our PDE (\ref{SCE}) now simplifies to  
\be 
( p_x \pa_x  +  p_y \pa_y ) \sbiC = 0   \m .  
\ee
By the flow method, this is equivalent to the ODE system 
\be 
\dot{x} =  p_x        \m , 
\ee
\be 
\dot{y} =  p_y        \m , 
\ee
\be 
\dot{p}_x = 0         \m , 
\ee
\be 
\dot{p}_y = 0         \m , 
\ee
\be 
\dot{\sbiC} = 0        \m , 
\ee
to be treated as a Free Characteristic Problem.  
Integrating the third and fourth equations  
\be 
p_x = u               \m , 
\ee
\be 
p_y = v               \m , 
\ee
so the first and second equations become 
\be 
\dot{x} = u           \m , 
\ee 
\be 
\dot{y} = v           \m , 
\ee
integrating in turn to 
\be 
x = u \, t + w        \m , 
\label{xut}
\ee 
\be 
y = v \, t            \m . 
\label{yvt}
\ee 
The final equation integrates to 
\be 
\sbiC  \es  \sbiC(u, \, v, \, w)  \m . 
\label{Cuvt}
\ee 
Eliminating $t$ between (\ref{yvt}) and (\ref{xut}) gives 
\be 
w  \es  x - \frac{p_x}{p_y}y                                  \m , 
\ee 
by which (\ref{Cuvt}) becomes 
\be 
\sbiC  \es  \sbiC\left(x \, p_y - y \, p_x , \, p_x , \, p_y   \right)  \m . 
\ee 
\n But 
\be
{p_x}^2 + {p_y}^2
\ee 
functional dependence is not proper. 
So, passing to coordinates 
\be 
p_{\pm}^{(2)} \m = \m {p_x}^2 \pm {p_y}^2
\ee 
we arrive at the final form 
\be 
\sbiC  \es  \sbiC\left(x \, p_y - y \, p_x , \,{p_x}^2 - {p_y}^2\right)         \m . 
\ee
These form the {\it function space of strong Chronos observables}   
\be 
{\cal O}_{\tiC}(1, \, 2, \, id, \, k)  \es  {\cal C}^{\infty}(\mathbb{R}^2)  \m .  
\ee

\subsection{\biN $\geq$ 3 in 1-$d$}

Our PDE (\ref{SCE}) having no further salient features with increasing $N$, we next pass to the general-$N$ solution in 1-$d$ 
By the flow method, this is equivalent to the ODE system 
\be 
\dot{q}^I    =  p^I                             \m , 
\ee
\be 
\dot{p}_I    =  0                               \m , 
\ee
\be 
\dot{\sbiC}  =  0                               \m , 
\ee
to be treated as a Free Characteristic Problem.  
Integrating the second vectorial equation,   
\be 
p_I = u_I                                       \m , 
\ee
so the first vectorial equation becomes 
\be 
\dot{q}^I = u^I                                 \m , 
\ee 
integrating in turn -- splitting off one component -- to 
\be 
q^i = u^i \, t + v^i                            \m , 
\label{qut}
\ee 
\be 
q^N = u^N \, t                                  \m . 
\label{quN}
\ee 
The final equation integrates to 
\be 
\sbiC = \sbiC(u^I, \, v^i)                      \m . 
\label{Cuv}
\ee 
Eliminating $t$ between (\ref{qut}) and (\ref{quN}) yields  
\be 
v^i  \es  q^i - \frac{p_i}{p_N}q^N              \m , 
\ee 
by which (\ref{Cuv}) becomes 
\be 
\sbiC  \es  \sbiC(q^i p_N - p_i q^N , \, p_I)   \m . 
\ee 
\n But 

\n\be
\sum_{I = 1}^N  {p_I}^2 
\ee 
functional dependence is not proper. 
So, passing to coordinates 

\n\be 
P^{(2)}    \:=  \sum_{I = 1}^N{p_I}^2                                \m , 
\ee 
\be 
P_i^{(2)}   :=  {p_i}^2 - {p_N}^2                                    \m ,
\ee 
we arrive at the final form 
\be 
\sbiC  \es  \sbiC(q^i p_N - p_i q^N , \,{p_i}^2 - {p_N}^2)           \m . 
\ee
These form 
\be 
{\cal O}_{\tiC}(1, \, N, \, id, \, k)  \es  {\cal C}^{\infty}(\mathbb{R}^{2 \, n}) \m .  
\ee

\subsection{The general case}

The $(d, N)$ version proceeds similarly, modulo making special use of just single component, without loss of generality $q^{N \, d}$, $p_{N \, d}$, giving
\be 
\sbiC  \es  \sbiC \left( q^{\Sigma} p_{N \, d} - p_{\Sigma} q^{N \, d} , \,{p_{\Sigma}}^2 - {p_{N \, d}}^2 \right)             \m ,
\ee
where $\Sigma$ runs from $1$ to $N \, d - 1$.

\m 

\n These form 
\be 
{\cal O}_{\tiC}(d, \, N, \, id, \, k)  \es  {\cal C}^{\infty}(\mathbb{R}^{2 ( N \, d - 1)}) \m .  
\ee

\section{Weak Chronos observables}

The {\it weak Chronos observables equation} is

\n\be 
- \bip \circ \bnabla \, \sbiC  \es  m \, \Chronos      \m \m \Rightarrow \m \m                             
- \sum_{I = 1}^N p_I \pa_{q^I} \sbiC  \es  m \, \left(  \frac{1}{2} \sum_{I = 1}^N  {p_I}^2 - W  \right)  \m : 
\label{WCE}
\ee  
an inhomogeneous extension of the strong Chronos PDE (\ref{SCE}). 
Thus the previous section has already provided the corresponding C.F.

\subsection{(\bid, \biN) = (1, 1)}

\n In this case, our PDE (\ref{WCE}) simplifies to 
\be 
- p \, \pa_x \sbiC   \es  m \left( \frac{p^2}{2} - W \right)        \m .
\label{2a}
\ee
While Sec 5.2 established that this does not support a proper C.F.,  this case does however already support a P.I.,  
\be 
\sbiC  \es  - m \, \frac{x}{p}  \left( \frac{p^2}{2} - W \right)
       \es  - m \, \frac{x}{p}              \, \Chronos             \m . 
\label{WCPI-11}	   
\ee
{\u{Derivation}} By the flow method, our P.D.E. is equivalent O.D.E. system 
\be 
\dot{x} = p                                                         \m , 
\ee 
\be 
\dot{p} = 0                                                         \m ,
\ee 
\be 
\dot{\sbiC} =  - m \left( \frac{p^2}{2} - W \right)                 \m . 
\ee 
by integrating its second equation to 
\be 
p = u        \m , 
\ee
so is first equation reads 
\be 
\dot{x} = u  \m , 
\ee 
immediately integrating to 
\be 
x = p \, t   \m , 
\ee 
so 
\be 
t  \es  \frac{x}{p}                                                 \m . 
\label{t = x/p}
\ee 
The last ODE is 
\be 
\dot{\sbiC} =  - m \left( \frac{u^2}{2} - W \right)                 \m , 
\ee 
immediately integrating to 
\be
\sbiC =  - m \left( \frac{p^2}{2} - W \right)t + \alpha 
\ee 
for constant $\alpha$ (but strike this out for not being proper), so by (\ref{t = x/p}), we arrive at our claimed P.I. (\ref{WCPI-11}).  $\Box$

\m 

\n{\bf Remark 1} This is not solely a function of constraints, so it is proper.
It is additionally a smooth function provided that $p \neq 0$ (this renders it a {\it local} weak Chronos observable `everywhere else').   
$k$ has the status of a free nonzero number: $k \, \in \mathbb{R}_*$,  so 
\be 
{\cal O}^w_{\tiC}(1, \, 1, \, Tr)  \es  \mathbb{R}_*        \m . 
\ee 
This is also of measure zero as compared to nontrivial strong observables spaces, a result extending to the general $(d, N)$ case.

\m	

\n (\ref{WCPI-11}) are not moreover momentum observables, so 
\be 
{\cal O}^w_{\tiC(\sbiP)}(1, \, 1, \, Tr) = \emptyset        \m . 
\ee 
%
%
%
		
\subsection{(\bid, \biN) = (1, 2)}

Here our PDE (\ref{WCE}) collapses to  
\be 
- ( p_x \pa_x + p_y \pa_y ) \sbiC   \es   m \left( \frac{{p_x}^2 + {p_y}^2}{2} - W \right)  \m . 
\label{2b}
\ee
By the flow method, this is equivalent to the ODE system 
\be 
\dot{x} = p_x     \m ,
\ee 
\be 
\dot{y} = p_y     \m , 
\ee
\be 
\dot{p}_x = 0     \m , 
\ee
\be 
\dot{p}_y = 0     \m , 
\ee
\be 
\dot{\sbiC}  \es  - m \left( \frac{{p_x}^2 + {p_y}^2}{2} - W \right) 
\ee 
Proceeding as in the previous subsection, we arrive at the same C.F. there, alongside the P.I. 
\be 
\sbiC  \es  - \left\{ a \, \frac{x}{p_x} + b \, \frac{y}{p_y}  \right\}  \left( \frac{{p_x}^2 + {p_y}^2}{2} - W \right) \mma a + b = m  
\ee  
using a general linear combination parallelling Sec \ref{Sym}'s.  

\m 

\n{\bf Remark 2} $m$ takes $\mathbb{R}_*$, but the P.I. takes values in all of $\mathbb{R}$.  
The P.I. is smooth except at $p_x = 0$ and $p_y = 0$, so this weak observable has a local character (c.f.\ the previous subsection). 
These form
\be 
{\cal O}^w_{\tiC}(1, \, 2, \, Tr)  \es  \mathbb{R} \times \mathbb{R}_*               \m . 
\ee

\subsection{The general case}

For general $(d, \, N)$, 

\n\be 
\sbiC                                            \es  - \sum_{a = 1}^{d} \sum_{I = 1}^{N}   \lambda_{Ia}  \frac{x^{Ia}}{p_{Ia}} \left( \frac{1}{2}\bip^2 - W \right)   \mma
\sum_{a = 1}^{d} \sum_{I = 1}^{N}  \lambda_{Ia}  \es   m
\ee 
solves likewise, forming 
\be 
{\cal O}^w_{\tiC}(d, \, N, \, Tr)  \es  \mathbb{R}^{N \, d - 1} \times \mathbb{R}_*        \m . 
\ee
It is also clear that our family of proper weak gauge observables does not qualify as weak momentum observables, so 
\be 
{\cal O}^w_{\tiC(\sbiP)}(d, \, N, \, Tr)  \es  \emptyset         \m . 
\ee

\section{Strong Dirac observables for \biV = const}

\subsection{r-approach}

If $Tr(1)$'s constraint $\scP$ is eliminated or never present prior to Assigning Observables by Taking Function Spaces Thereover, the remaining `energy' constraint 
induces the {\it strong Dirac observables brackets equation} 
\be
\mbox{\bf \{} \w{\Chronos} \mbox{\bf ,} \, \sbiD \mbox{\bf \}}  \es  0  \m , 
\label{R-D-BE} 
\ee
in mass-weighted relative Jacobi coordinates.
This implies the {\it strong Dirac observables PDE} 

\n\be 
\sum_{i = 1}^n \pi_i \pa_{\rho^i} \w{\sbiD}  \es  0 . 
\ee 
But this is just equivalent the $(1, \, n)$ Chronos problem, so we can immediately write down its solution, 
\be 
\w{\sbiD}  \es  \w{\sbiD}\left(\rho^{\sigma} \pi_N - \rho^N \pi_{\sigma} , \, {\pi_{\sigma}}^2 - {\pi_N}^2\right)                    \m ,
\ee 
forming the function space of observables 
\be 
{\cal O}_{\w{\tiD}}(1, \, N, \, Tr, \, k) \es  {\cal C}^{\infty}\left(\mathbb{R}^{2 ( n \, d  - 1)}\right)  \m .  
\ee  
This case does not support configuration observables by argument (\ref{Res-Empty}), but it does support momentum observables, 
\be 
\w{\sbiD}  \es  \w{\sbiD}({\pi_{\sigma}}^2 - {\pi_N}^2)                                                     \m . 
\ee 
forming 
\be 
{\cal O}_{\w{\tiD}(\sbiP)}(d, \, N, \, Tr, \, k) \es {\cal C}^{\infty}\left(\mathbb{R}^{n \, d - 1}\right)  \m .  
\ee 
\n{\bf Remark 1} $(d, \, N) = (1, \, 3)$ is minimal for these function spaces to be nonempty; in this case, 
\be 
\w{\sbiD}  \es  \w{\sbiD} \left( \rho^1 \pi_2 - \rho^2 \pi_1 , \, {\pi_1}^2 - {\pi_2}^2 \right)            \m ,
\ee 
forming the function space of observables 
\be 
{\cal O}^w_{\w{\tiD}}(1, \, 3, \, Tr, \, k) \es {\cal C}^{\infty}(\mathbb{R})
\ee  
and admitting restriction to the momentum observables 
\be 
\w{\sbiD}  \es  \w{\sbiD}({\pi_1}^2 - {\pi_2}^2)                                             \m ,  
\ee 
themselves forming 
\be 
{\cal O}^w_{\w{\tiD}(\sbiP)}(1, \, 3, \, Tr, \, k) \es {\cal C}^{\infty}(\mathbb{R})  \m .  
\ee

\subsection{Indirect approach's equations}

Models with gauge group of physically meaningless transformations $Tr(d)$, in possession also of a Chronos constraint reinterpretation of the energy equation, 
support furthermore a notion of Dirac observables, which, in the strong case, obey the {\it strong Dirac observables brackets equations},

\n\be 
\mbox{\bf \{} \u{\scP}    \mbox{\bf ,} \,  \sbiD \mbox{\bf \}}  \es  0  \m , 
\label{D-BE-1} 
\ee  
\be 
\mbox{\bf \{} \Chronos \mbox{\bf ,} \,   \sbiD \mbox{\bf \}}  \es  0  \m .
\label{D-BE-2}
\ee
This implies the {\it indirect formulations strong Dirac observables PDE system} 

\n\be 
\sum_{I = 1}^N \nabla_{\u{q}^I} \sbiD \es 0                            \m , 
\label{SDE-1}
\ee 

\n\be 
\bp \circ \bnabla _, \sbiD  \es  \sum_{I = 1}^N p_I \pa_{q^I} \sbiD 
                            \es                 0                      \m . 
\label{SDE-2}
\ee

\subsection{Indirect approach for (\bid, \biN) = (1, \, 1)}

In this case, our system collapses to
\be
\pa_x \sbiD = 0        \m , 
\ee 
\be 
p \, \pa_x \sbiD = 0   \m .  
\label{2-Fac}
\ee
Now if $p = 0$ is used to solve the second equation, the first equation's 
\be 
\sbiD = \sbiD(p) 
\ee 
collapses to just a constant (so it goes from not being proper in second sense to not being proper in the first sense). 
But if the second factor of (\ref{2-Fac})'s PDE is adopted, this does not form a system with the first equation, but rather identically coincides with it.  
Thus solving the strong gauge-observables equation suffices, but we have already shown this to yield no proper observables. 
Thus 
\be 
{\cal O}_{\tiD}(1, \, 1, \, Tr)  \es  \emptyset   \m . 
\ee 
\n{\bf Remark 1} Another way of seeing this is that neither gauge nor Chronos subsystems support any proper strong observables, 
and then applying argument (\ref{Res-Empty}) to either.

\subsection{Indirect approach for (\bid, \, \biN) = (1, \, 2), (1, \, 3) and the general case}\label{SD-12}

In this case, our system (\ref{SDE-1}, \ref{SDE-2}) simplifies to  
\be
( \pa_x + \pa_y ) \sbiD = 0                 \m , 
\ee 
\be 
(p_x \, \pa_x + p_y \, \pa_y) \sbiD  =  0   \m .  
\ee
\n The centre of mass sequential method \cite{PE-1} applies, further collapsing our system to the (1, 1) strong Chronos observables problem in $r$, $\pa_r$ variables, for 
\be 
r = x - y                                   \m ,  
\ee 
\be 
p_r = p_x - p_y                             \m .
\ee 
Thus we know from Sec \ref{SC-12} that there are no proper strong Dirac observables for $N = 2$: 
\be 
{\cal O}_{\w{D}}(1, \, 2, \, Tr, \, k)  \es  \emptyset  \m . 
\ee 
So we are  left with 
\be 
{\cal O}_D(1, \, 2, \, Tr, \, k)  \es  \emptyset  \m . 
\ee 
\n{\bf Remark 1} Solving (\ref{SDE-2}) piecemeal in the general case reveals dependence on differences. 
This moreover admits reformulation by linear combination, 
with $N = 3$ being minimal for nontrivial involvement of the mass-weighted Jacobi coordinates featuring in all subsequent cases.  
Then rephrasing the second equation in terms of this returns Sec 7.1's sole r-equation. 
Thus 
\be 
\widetilde{\sbiD}  \es  \sbiD 
\ee
for this model already holds at the level of strong observables equations, and thus continues to hold at the level of solutions and of function spaces that these form.  
In this way, Sec 7.1's workings suffice to characterize the outcome of the general $(N, \, d)$ case's indirect approach as well.

\section{Weak Dirac observables}

\subsection{Indirect approach's equations}

The corresponding {\it weak Dirac observables equation system} is  

\n\be 
-\sum_{I = 1}^N \nabla_{\u{q}^I}     \sbiD  \es  k \, \u{\scP}                   \m + \m  \u{l} \, \Chronos  
                                            \es  k \, \sum_{I = 1}^N \u{p}_I  \m + \m  \u{l} \left( \sum_{I = 1}^N \frac{{p_I}^2}{2} - W \right)                   \m , 
\label{WD-1}
\ee 

\n\be 
- \bip \circ \bnabla \siD  \es  \u{m} \cdot \u{\scP} \m + \m n \, \Chronos \m \m \Rightarrow \m \m 
- \sum_{I = 1}^N \u{p}_I \cdot \nabla_{\u{q}^I} \sbiD  \es  \u{m} \cdot \sum_{I = 1}^N \u{p}_I  \m  + \m  n \left( \sum_{I = 1}^N \frac{|p_I|^2}{2} - W \right)   \m . 
\label{WD-2}
\ee 
This is an inhomogeneous extension of PDE system (\ref{D-BE-1}, \ref{D-BE-2}).

\subsection{(\bid, \, \biN) = (1, \, 1)}

Here our PDE system (\ref{WD-2}, \ref{WD-2}) collapses to 
\be 
- \pa_x \sbiD    \es  j  \m + \m  k \, p  \m + \m  l \, p^2    \m . 
\label{WD-1-1}
\ee 
\be 
- p \pa_x \sbiD  \es   o  \m + \m  m \, p \m + \m  n \,p^2     \m , 
\label{WD-1-2}
\ee
for free constants $k$, $l$, $m$, $n$ and $j := - l \, W$, $o :=  - n \, W$.

\m 

\n Let us first consider the free zero-energy problem, for which $W = 0$ and $j = 0 = o$.  
Now substituting (\ref{WD-1-1}) in (\ref{WD-1-2}) yields
\be 
p( l \, p^2 + (n - k)p + m )  = 0                   \m .  
\ee 
So $p = 0$ (but this returns us to the strong gauge observables problem) or 
\be 
p = \frac{k - n \pm \sqrt{(k - n)^2 - 4 m l}}{2 l}  \m .  
\ee 
Define the discriminant 
\be 
D := (k - n)^2 - 4 m l   \m .  
\ee 
This gives 2 fixed $p$-value solutions for $D > 0$,  
a single     fixed $p$-value solution  for $D = 0$,  
or no solutions at all                 for $D < 0$ 
(since $p$ is a physical quantity and so must take a real value).  
We do not moreover consider these fixed-$p$ solutions to be particularly interesting as regards the development of the general theory of observables. 

\m 

\n Next bringing in constant $W$, substituting (\ref{WD-1-1}) in (\ref{WD-1-2}) yields
\be 
l \, p^3  \m + \m  (k - n) \, p^2  \m +  \m (j - m) p - o  \es  0  \m .  
\ee 
This less concisely returns 1, 2 or 3 fixed values of $p$ (by Cardano's formula, and noting that cubics always have at least one real root).  

\m 

\n Thus weak Dirac observables are only supported by this model for `special balancing values' of $p$; 
in other words, $N > 1$ is required before such models can start to return a standard theory of observables. 

\m 

\n{\bf Remark 1} Weak Dirac observables are not just a restriction of weak-Chronos by weak-$Tr$ or vice versa. 
This is because in the weak Dirac system, weak Chronos contains a $Tr$-constraint $\u{\scP}$ piece and weak-$Tr$ contains a $\Chronos$ constraint piece. 
Because of this, weak observables systems do not involve as simple a restriction hierarchy as strong observables ones do. 

\m

\n{\bf Remark 2} $\lFrg$-weak observables are coupled only one way round: $\lFrg$-constraint term in Chronos equation but no Chronos constraint term in the $\lFrg$-equation.  
This corresponds in our example to the $l = 0 = n$ case.  
The $\lFrg$-weak system contains the weak $\lFrg$-observables equation, so $\lFrg$-weak Dirac observables can be viewed as a restriction of $\lFrg$-weak $\lFrg$-observables.  

\m

\n This is to be understood as a finer distinction, in some ways comparable with Dirac providing \cite{Dirac} primed and starred Hamiltonians 
as well as bare and total Hamiltonians.

\subsection{(\bid, \, \biN) = (1, \, 2) indirect approch}

Our PDE system (\ref{WD-1}, \ref{WD-1}) now collapses to 
\be 
- ( \pa_x + \pa_y ) \sbiD         \es  k ( p_x + p_y )  \m + \m  l \left( \frac{ {p_x}^2 + {p_y}^2 }{2} - W \right)    \m , 
\ee 
\be 
- ( p_x \pa_x + p_y p_y ) \sbiD   \es  l ( p_x + p_y )  \m + \m  n \left( \frac{ {p_x}^2 + {p_y}^2 }{2} - W \right)    \m , 
\label{2d}
\ee
for free constants $k$, $l$, $m$ and $n$.

\m 

\n We know there is no C.F. from Sec \ref{SD-12}, and that the first equation's first P.I. from Sec \ref{Sym}.
The second P.I. is 
\be 
\sbiD  \es  - l \, \left( \frac{ {p_x}^2 + {p_y}^2 }{2} - W  \right)  t + \gamma       \m , 
\ee 
but we strike out the $\gamma$ for not being proper.
We then use the strong Chronos working to remove $t$, arriving at  
\be
\sbiD  \es  -    ( c \, x + d \, y  )  \left( \frac{ {p_x}^2 + {p_y}^2 }{2} - W \right)    \mma 
c + d = l                                                                             \m .
\ee
in the general linear combination case. 
We next substitute the sum of our 2 P.I.'s into the second equation. 
This gives 
\be 
a \, {p_x}^3 + b \, {p_y}^3 + a \, p_x{p_y}^2 + b \, p_y{p_x}^2            + 
(2 \, c + n) {p_x}^2 + (2 \, d + n) {p_y}^2                                          + 
2(m - \, W )( p_x + p_y ) - 2 \, n \,  W  =  0           \m .  
\ee
For free $p_x$, $p_y$, this forces $k = l = m = n = 0 = W$. 
Thus we are left with a cubic in 2 variables to solve for fixed $p_x$, $p_y$.

\m 

\n This shows that 
\be 
{\cal O}^w_{\w{\tiO}} \m \neq \m  {\cal O}^w_{\tiO}
\ee 
is possible, by $\sbiO$ having more room for P.I.s than $\w{\sbiO}$ does.

\subsection{Reduced approach's equations}

The corresponding {\it weak reduced Dirac observables equation system} is  

\n\be 
- \sum_{I = 1}^N \u{\pi}_I \cdot \nabla_{\u{\rho}^i} \sbiD  \es  n \left( \sum_{I = 1}^N \frac{|\pi_i|^2}{2} - W \right)   \m . 
\label{R-SDE} 
\ee

\subsection{Reduced approach for (\bid, \, \biN) = (1, \, 2)}

In this case, our PDE (\ref{R-SDE}) simplifies to  
\be 
- ( p \, \pa_r ) \w{\sbiD}  \m = \m  o  \m + \m  m \, {p_r}^2                                                   \m , 
\label{2c}
\ee
This is of the same form as the $\scC(1, \, 1)$ problem, so is solved by 
\be 
\w{\sbiD}  \es  - m \, \frac{r}{p_r} \left( \frac{{p_r}^2}{2} - W \right) 
           \es  - m \, \frac{x - y}{p_x - p_y} \left( \frac{{p_x - p_y}^2}{2} - W \right)       \m .
\ee
\be 
0  =  p_r  
   =  p_x - p_y 
\ee 
is now the form of the case that locally excepts smoothness.
Thus 
\be 
{\cal O}_{\w{\tiD}}^w(1, \, 2, \, Tr, \, k)   \es   \mathbb{R}  \m .
\ee

\subsection{The general reduced case}

We finally present the general $(d, \, N)$ case, in the more interesting reduced version.

\n\be 
\sbiD                                     \es  - \sum_{\delta = 1}^{n \, d} \lambda_{\delta} \, \frac{\rho^{\delta}}{p_{\delta}} \left( \frac{1}{2}\bipi^2 - W \right) \mma  
\sum_{\delta = 1}^{n \, d} \lambda_{\delta}  \es  m 
\ee 
solves, forming 
\be 
{\cal O}^w_{\tiD}(d, \, N, \, Tr)  \es  \mathbb{R}^{n \, d - 1} \times  \mathbb{R}_*        \m . 
\ee
It is also clear that our single $m$-family of proper weak Dirac observables does not qualify as weak momentum observables, so 
\be 
{\cal O}^w_{\tiD(\sbiP)}(d, \, N, \, Tr)  \es  \emptyset         \m . 
\ee

\vspace{10in}

\section{Conclusion}

\subsection{Summary of results}

We have considered observables equations, solution of which gives systematically a given physical theory's classical canonical observables.
Ab initio, these equations take the form of zero Lie brackets with the sum-over-points of generators. 
We moreover recast these observables equations as PDEs, and find the corresponding solution spaces. 
\n The current article and \cite{PE-1} represent the first major extension of \cite{ABook}'s treatment of Background Independence and the Problem of Time, 
by providing a concrete theory of solving for observables on the endpoint of the first branch of Fig 1.b) for now in the local, classical finite-theory setting.  
We have moreover now covered the following types of diversity among canonical observables by justifying each concept, providing specific PDEs for each concept, 
and providing concrete examples for many of the possible combinations of these concepts as supported by a simple model: translational- and reparametization-invariant mechanics.  

\m

\n 1) Strong, properly-weak and weak notions of observables, defined within Dirac's conceptualization at the level of brackets equations, 
with properly-weak meaning those weak observables which are not already strong observables. 
We showed that these notions are moreover implemented at the level of linear PDE systems by the complementary function, particular integral and complete solution respectively. 
We also clarified how for sufficiently large $N$ to have nontrivial degrees of freedom, properly-weak observables are but a small (measure-0) extension of strong ones. 
This is due to the complementary function admitting free characteristic treatment thus giving arbitrary sufficiently-smooth functions of characteristic coordinates, 
whereas the particular integral at most depends on the values of a finite set of constants.

\m 

\n 2) Indirect and r-formulations' notions of observables, where `r' stands for reduced, or, as a coincident notion for our models, 
directly configurationally relationally formulated.
In particular, we exemplified how indirect formulations, through having more constraints, 
have more room both for examples of properly-weak observables and for conceptual subtypes thereof. 

\m

\n 3) Restrictions of each full notion of observable to pure-momentum and pure-configuration observables, 
the latter coinciding with purely geometrical preserved quantities for the corresponding notion of space's geometry.  

\m 

\n 4) There is a distinct notion of observables for each closed subalgebraic structure of constraints.
The model in hand  supports four such subalgebras of note: the trivial one, the translational constraint $\u{\scP}$ by itself, the $\Chronos$ constraint by itself, 
and the full constraint algebra.
These form the square lattice at the bottom of Fig 3.b).  
The quotient spaces form the dual lattice, and the lattice of notions of observables is on this as well, and thus is a dual lattice too. 
These lattices are presented further up in Fig 3.b). 
3.a) and 3.c) are the momentum and configurational observables counterparts of these respectively.  

\m

\n Our simple model turns out to support an extensive, if not complete, repertoire of the $3 \times 2 \times 3 \times 4 = 72$ combinations of notions 1) to 4), 
some being trivial for this model and some coinciding with each other as per Fig 3.a-c).  
Each of the 22 realized notions of observables' function space of observables is summarized in Figs 3.d-f) in presheaf form.   
%
{            \begin{figure}[!ht]
\centering
\includegraphics[width=1.0\textwidth]{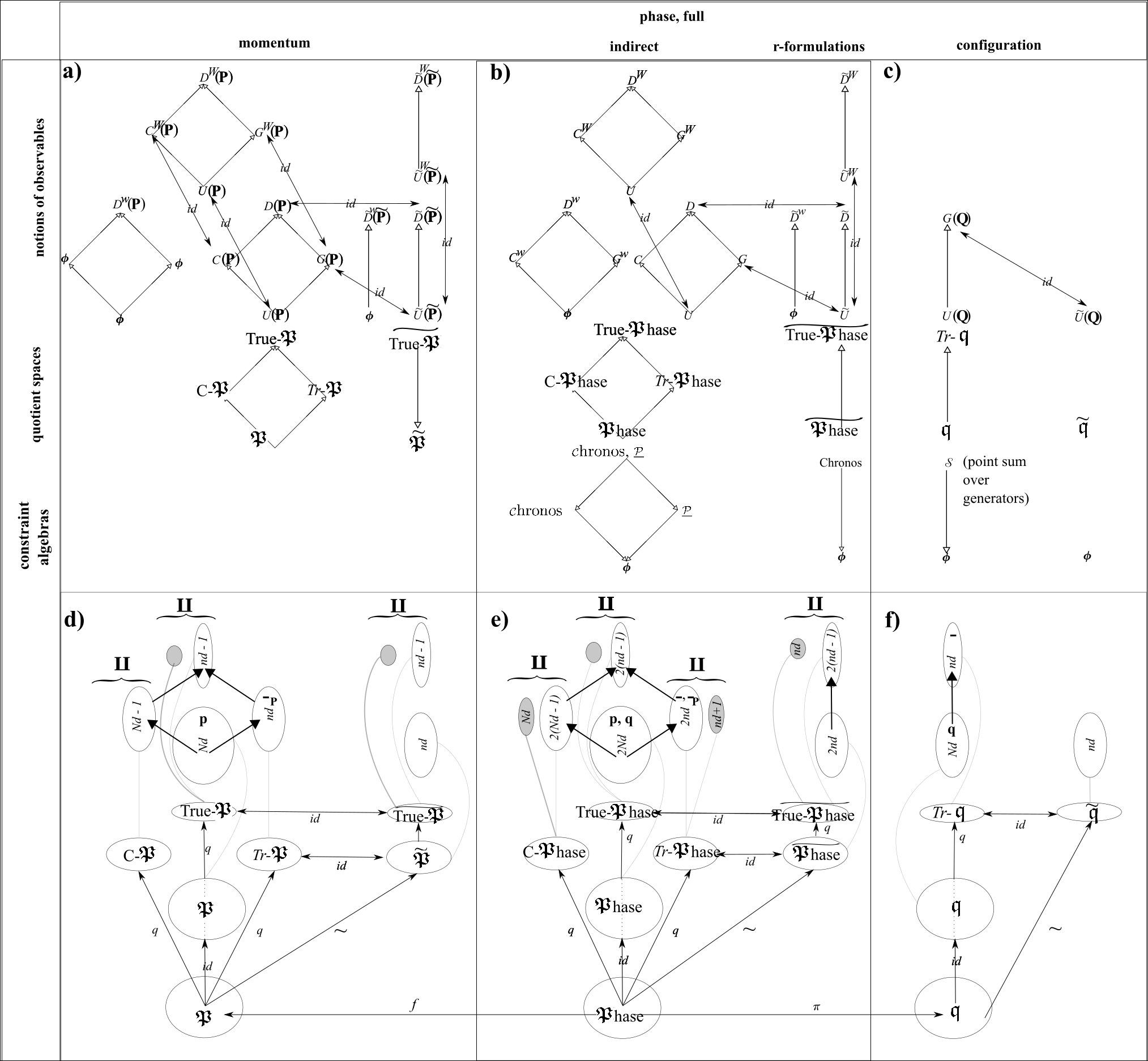}
\caption[Text der im Bilderverzeichnis auftaucht]{\footnotesize{Lattice and presheaf summary 
of the current Article's notions of observables and their realization as function spaces.
White arrows denote order relations, 
$q$ quotient map, 
$\pi$ projection onto the base space, 
$f$ the picking out of a fibre, 
$\sim$ reduction, and 
black flat-backed arrows denote presheaf restriction maps. 
In the upper layer of d) to f), white shading indicates strong observables, whose labels denote the dimension $a$ of ${\cal C}^{\infty}(\mathbb{R}^a)$, 
with thin grey lines denoting strong observables presheaf functors ${\cal F}$. 
Grey shading indicates properly-weak observables, whose labels denote the dimension $b$ of the much smaller $\mathbb{R}^{a- 1}\times \mathbb{R}_*$, 
with thick grey lines denoting properly-weak observables presheaf functors ${\cal F}^w$.
Shorthands for the simpler functional dependences of the strong observables are also indicated in bold. 
    Everything in e) has furthermore a restriction map $\mbox{res}^{\tPhase, \tFrP}$ leading into its counterpart in d), 
and everything in f)                 a restriction map $\mbox{res}^{\tPhase, \tFrQ}$ leading from its counterpart in e).}}
\label{Presh} \end{figure}          }

\subsection{Further research directions}

\n{\bf Frontier A} Article 2 considers observables PDEs and their solutions for mechanics \cite{FileR, AMech} on Kendall's preshape space \cite{Kendall84, Kendall}, 
i.e.\ quotienting out dilations as well as translations and thus overall quotienting out dilatations.  
Article 3 furthermore considers these matters for mechanics \cite{B03, FORD, FileR} on Kendall's shape space \cite{Kendall84, Kendall}, 
i.e.\ quotienting out rotations as well, and thus overall quotienting out similarities. 
This also permits consideration of the subcase in which just translations and rotations -- i.e.\ Euclidean transformations -- are quotiented out, 
which corresponds to the most traditional setting \cite{BB82} of the relational side of the Absolute versus Relational Motion Debate \cite{L, M, Buckets, ABook}.  

\m 

\n There is moreover a more technical description of such extensions which is of particular use upon entertaining further affine, projective, conformal and supersymmetric 
generalization \cite{Bhatta, PE16} of Kendall's Shape Theory and of Mechanics thereupon \cite{AMech, ABook}. 
Namely that, firstly, the translational constraint is a homogeneous-linear function of the momenta alone. 
Secondly, dilational, rotational and affine constraints are homogeneous-bilinear in configurations and momenta \cite{AMech, ABook}. 
Thirdly, special-projective and special-conformal constraints are homogeneous-linear in momenta and homogeneous quadratic in configurations \cite{PE-3, Brackets-I, Brackets-II}. 
(Supersymmetric constraints are also homogeneous-bilinear, albeit now partially through involvement of Grassmann variables \cite{AMech}).  
This gives the following laddered pattern.  

\m 

\n{\bf Rung I)} Constraints homogeneous-linear in the momenta alone experience the current Article's loss of pure-momentum notions of constraint. 

\m 

\n{\bf Rung II)} The homogeneous bilinear constraints that collectively correspond to the general linear group 
-- dilations, rotations and the furtherly affine shears and Procrustes stretches -- exhibit instead symmetry between the forms taken by configuration and momentum observables. 

\m  

\n{\bf Rung III)} Constraints homogeneous-linear in momenta and homogeneous-quadratic in configurations 
are the first to have nontrivial and non-symmetrically related momentum observables.

\m  

\n{\bf Rung IV)} Constraints that are cubic-order or higher in the configurations fail to close as a finite algebra. 
By this, shape theory based on quotienting finite constellation spaces by such a group trivializes due to no degrees of freedom surviving. 
So higher polynomiality constraints (or underlying generators) fails to produce further shape theories \cite{Brackets-II}.  

\m 

\n Each rung represents a considerable step-up in complexity, so it makes very good sense to start with translations, and then extend by general-linear transformations, 
of which Articles 2 and 3 are a nested subcase. 
\cite{PE-1, PE-2, PE-3} covered this progression in the simpler case of geometrical preserved quantities. 
The full-observables case, however, has much further notional diversity as well as the following two further complications. 

\m 

\n{\bf Complication 1}: need for new physical actions. 
Beyond the case of Kendall's shape space and the similarity Mechanics thereupon, 
one needs to supply actions whose kinetic matrices are built to be affine, projective, conformal and/or supersymmetry invariant.
While \cite{AMech} provided affine, conformal and supersymmetric such, it only provided the simplest 1-$d$ supersymmetric action and no projective action.
These two complications combine to render the full observables analysis slower-paced than the full preserved quantities analysis.  

\m 

\n{\bf Complication 2}: inclusion of nontrivial potentials. 
As explained in \cite{FileR}, our take on this is to a priori prescribe potentials that are invariant under the geometrical automorphism group in question, 
parallelling the kinetic terms used in this regard. 
These do not spoil the constraint algebra (while other potentials, rather predictably, do spoil it).
In the current Article's case of translations, this means using e.g.\ 
\be 
V = V(||\biq|| \m \mbox{alone}) \m 
\label{V1}
\ee
or its generalization 
\be 
V = V(\biq \cdot \biq  \m \mbox{alone})   \m 
\label{V2}
\ee
is appropriate and indeed consistent. 
One can also generalize by 
\be 
V = V(||\biq|| , \, ||\biq \pm \biq|| \m \mbox{alone}) \m , 
\label{V3}
\ee
which is equivalent to the preceding by the polarization identity. 
Once one has nontrivial potential functions, however, every notion of observables corresponding to a subalgebra containing $\Chronos$ 
requires explicit solution afresh for each choice of function $V$, giving an infinity of calculations (almost none of which permit exact solutions).  
There is moreover no counterpart of this complication in evaluating geometrical preserved quantities, or concrete examples of notions of observables corresponding to subalgebras 
not containing $\Chronos$.

\m 

\n{\bf Frontier B} Obtaining the preserved quantities for each geometry as the solution of a concrete PDE \cite{PE-1, PE-2, PE-3} 
is clearly a very small corner of the above observables program.
This is none the less of considerable interest \cite{8-Pillars} since it provides new Foundations of Geometry \cite{Hilb-Ax, HC32, Coxeter, S04, Stillwell}, 
and moreover ones with a sharp systematic tool for computing invariant quantities.
This is to be compared with the generalized Killing equation \cite{Yano}: a sharp systematic PDE tool for computing automorphism groups. 

\m 

\n{\bf Frontier C} Our much expanded theory of notions of observables applies moreover to {\sl whichever other} kind of physical model, 
and our concrete methods of solution apply furthermore to whichever other finite classical model.  
The often-used minisuperspace models do not however exhibit many of the current article's variety of notions of observables 
due to having but a single scalar non-linear constraint, by which 

\n\be 
\sbiU = \sbiG = \sbiK \m \mbox{ and } \m 
\sbiC = \sbiD \m . 
\ee
There is also rather less scope for different notions of weak constraints to attain distinct realizations in such a setting.  

\m 

\n There are also the Introduction's Ambiguity 0-or-1 and Complexities 1, 2 and 3 to contend with: spacetime versus canonical notions, constraints, quantization and global treatment. 
All combinations of Complexities 1, 2, 3 can be applied to either of 0-or-1, 
whereas the current article only considers concrete equations and their solutions for Complexity 1's constraints in the local classical canonical setting. 
It is also worth noting that the quantum version of the local resolution of the Problem of Time in \cite{ABook} requires quantum unconstrained observables up front 
-- the so-called kinematical quantization \cite{I84} -- i.e.\ prior to addressing Facets 1, 2 and 3. 
Thereafter, one needs to adjust one's unconstrained observables to quantum constrained observables, as the `harder half' of Facet 4's Assigning Observables.  
In this way, Assigning Observables undergoes a further quantum-level correction split beyond the depiction of order of addressal of Facets in Fig 1.b).

\m

\n There is moreover a further Complexity 0 to take into account once one is formulating observables equations as DEs. 

\m 

\n{\bf Complexity 0} Field Theories's observables DE's are moreover not partial but {\it functional} DEs (FDEs). 
As compared to finite theories' PDEs, rather less is known about FDEs in general, or even for our concrete case of interest: 
(systems of) linear FDEs to be treated as Free Characteristic Problems.  
This makes for a good mid-term focus for follow-up papers extending the current Series' research.

\m 

\n{\bf Complexity 0'} Some prominent Field Theories which are additionally Gravitational Theories, such as GR and Supergravity, 
exhibit a further complexity at the underlying level of brackets equations. 
Namely their constraint algebraic structures are not Lie algebras with their structure constants but rather much harder Lie algebroids \cite{Bojowald12, AMech, ABook, Bojowald17} 
with structure functions instead.  
This accounts for some `algebraic structures' phrasings in the current Article, this term being taken to be a portemanteau of Lie algebra and Lie algebroid cases.  

\m 

\n All in all, the current Article represents but the local classical constrained canonical finite `tip' of the canonical observables `iceberg'. 
This is spanned by the field-theoretic, constrained, quantum and global complexities, and there is a comparable spacetime observables `iceberg' out there as well.

\m 

\n{\bf Frontier D}: {\it Comparative study of Background Independence} The current series of Artices is furthermore of relevance to this study of Background Independence 
level by level of structure. 
I.e.\, in more detail, a systematic study of a) geometrical-level Background Independence for each level of geometrical structure \cite{AMech, ABook}
(Euclidean, similarity, affine, projective, conformal, supersymmetric...).
b) Subsequent levels of mathematical structure \cite{I89-I91, ICat, ASoS, ABook} 
(e.g.\ topological manifolds, topological spaces, sets in the standard foundational system of mathematics, 
or of such as categories, sheaves and topoi in other foundational systems).  

\m 

\n As \cite{PE-1} outlines and \cite{5-6-7} describes, there is moreover interplay between Frontiers B and D.  
For some of the Foundations of Geometry (old and new) are simpler subcases or `shadows' of the approaches used in \cite{ABook}'s local resolution of the Problem of Time and 
(comparative) formulation of Background Independence. 
  
\m 

\n{\bf Acknowledgments} On this occasion, I thank all who encouraged my conceptual thinking.  
O, 
GJ, 
EBJ, 
JL, 
AA,
ER,  
L,
C, 
Malcolm MacCallum, 
Julian Barbour, 
Niall \'{o} Murchadha, 
Jimmy York, 
Jurgen Ehlers, 
Karel Kucha\v{r}, 
Bryce DeWitt, 
Reza Tavakol, 
Don Page,  
Enrique Alvarez, 
Marc Lachi$\grave{\me}$ze-Rey, 
and Jeremy Butterfield.  
Chris Isham most of all, 
the three A's 
and the many S's.


\end{document}